\documentclass{elsart}

\usepackage{graphicx}
\usepackage{epsfig}
\usepackage{booktabs}
\usepackage{multirow}
\usepackage{rotating}

\usepackage{amsmath, amssymb, amsfonts}
\usepackage{color}

\begin{document}

\begin{frontmatter}



\title{Impact of Charge Trapping on the Energy Resolution of Ge Detectors for Rare-Event Physics Searches}
\author[usd]{D.-M. Mei\corauthref{cor}}, 
\corauth[cor]{Corresponding author.}
\ead{Dongming.Mei@usd.edu}
\author[iserc]{R.B Mukund},
\author[usd]{W.-Z. Wei},
\author[usd]{R. Panth},
\author[usd]{J. Liu},
\author[usd]{H. Mei},
\author[usd]{Y.-Y. Li},
\author[usd]{P. Acharya},
\author[usd]{S. Bhattarai},
\author[usd]{K. Kooi},
\author[usd]{M.-S. Raut},
\author[yzu]{X.-S. Sun},
\author[usd] {A. Kirkvold},
\author[usd,yzu]{K.-M. Dong},
\author[usd]{X.-H. Meng},
\author[usd]{G.-J. Wang},
\author[usd]{G. Yang},

\address[usd]{Department of Physics, The University of South Dakota, Vermillion, South Dakota 57069, USA}
\address[iserc]{ISERC, Visva Bharati University, Santiniktean, West Bengal 731235, India}
\address[yzu]{School of Physics and Optoelectronic, Yangtze University, Jingzhou 434023, China}

\begin{abstract}
    Charge trapping degrades the energy resolution of germanium (Ge) detectors, which require to have increased experimental sensitivity in searching for dark matter and neutrinoless double-beta decay. We investigate the charge trapping processes utilizing nine planar detectors fabricated from USD-grown crystals with well-known net impurity levels. The charge collection efficiency as a function of charge trapping length is derived from the Shockley-Ramo theorem. Furthermore, we develop a model that correlates the energy resolution with the charge collection efficiency. This model is then applied to the experimental data. As a result, charge collection efficiency and charge trapping length are determined accordingly. Utilizing the Lax model (further developed by CDMS collaborators), the absolute impurity levels are determined for nine detectors. The knowledge of these parameters when combined with other traits such as the Fano factor serve as a reliable indicator of the intrinsic nature of charge trapping within the crystals. We demonstrate that electron trapping is more severe than hole trapping in a p-type detector and the charge collection efficiency depends on the absolute impurity level of the Ge crystal when an adequate bias voltage is applied to the detector. Negligible charge trapping is found when the absolute impurity level is less than 1.0$\times$10$^{11}/$cm$^{3}$ for collecting electrons and 2.0$\times$10$^{11}/$cm$^{3}$ for collecting holes. 
\end{abstract}
\end{frontmatter}

\maketitle

\section{Introduction}
\label{sec:intro}

One of the consequences the various extensions of the Standard Model of particle physics is that the neutrinos are their own anti-particles~\cite{1,2,3}. Currently, experiments have been set up or are under preparation to observe neutrinoless double-beta ($0$$\nu$$\beta\beta$) decay which would have the discovery potential to confirm the Majorana nature of the neutrinos~\cite{frank, alan}. The major distinguishing factor between these experiments is the choice of detector materials being used~\cite{4,5,6}. Experiments like GERDA~\cite{5} and M\textsc{ajorana} D\textsc{emonstrator}~\cite{6} utilize high-purity germanium (HPGe) detectors which have been enriched with $^{76}Ge$ to carry out their searches for $0$$\nu$$\beta\beta$ decay. The key parameter in all these experiments is the energy resolution of their spectra. HPGe detectors with the best energy resolution allow a sharp peak in the spectra within the region of interest (ROI) to be an indicator of a potential $0$$\nu$$\beta\beta$ decay. This implies that the intrinsic 2$\nu\beta\beta$ background is well separated from $0$$\nu$$\beta\beta$ in the ROI and other backgrounds are minimized using a narrow ROI due to excellent energy resolution. 

Another area of astroparticle physics which has seen considerable progress is the field of dark matter searches. The evidence of the existence of dark matter has been convincingly proven via various astronomical observations performed over the past few decades~\cite{9,10,11,12,13}. Those observations have shown that a majority of matter in the universe is dominated by dark matter(~85\%). Out of various hypothetical particles predicted by different models, the favoured candidates for most of the ongoing experiments are WIMPs (Weakly Interacting Massive Particles) which can interact very weakly with ordinary matter and be observed by direct observations~\cite{dama2017, cdms1,edel1, xenon10, xenon100, lux, pandax, xmass, zeplin, xenon1t, kim, cosine, supercdms, cdex, cea}, giving us an estimate of the effective mass of the incident WIMPs~\cite{16}.
As such, the detector energy response to these low-energy signals, which may be generated due to WIMP-nuclei collisions within a detector, is a crucial parameter that must be optimized for higher detector sensitivity. It has been shown that Ge detectors with excellent energy resolution can provide a reliable low-energy threshold for dark matter searches~\cite{cdms, edeweiss, cdex, cea}. In fact, the lowest energy threshold and the best discrimination ability in identifying nuclear recoils from electronic recoils is obtained by Ge detectors~\cite{cdms, edeweiss}.

Ge detectors collect charge created by the energy deposition from incoming particles interacting with Ge atoms through drifting charge carriers across the detector under an electric field. Since impurities exist in detector crystals~\cite{7}, the charge carriers will encounter them while drifting through the detector. A charge carrier can become bound to a spatially-localized impurity state and hence no longer contributes to the drift current signal temporarily or permanently. This process is known as charge trapping and the charge carrier is then said to be trapped or captured by the impurity atoms.

 Charge trapping contributes to the broadening of a Ge detector's energy resolution and it occurs due to existing impurities inside detector crystals~\cite{7}.  During the charge drifting process, charge carriers can be trapped by impurities resulting in either a prolonged charge pulse or a complete charge loss. The former one can be partially or fully recovered through charge trapping correction by measuring drift time~\cite{mls, ryan}. The latter is a capture process leading to a permanent charge loss. Although effective charge correction methods have also been developed for correcting the permanent charge loss in various Ge detectors~\cite{5,6}, understanding the physical mechanism of charge trapping is important to further improving charge collection efficiency and energy resolution.

\begin{figure}[htb!!!]
\includegraphics[angle=0,width=14.cm] {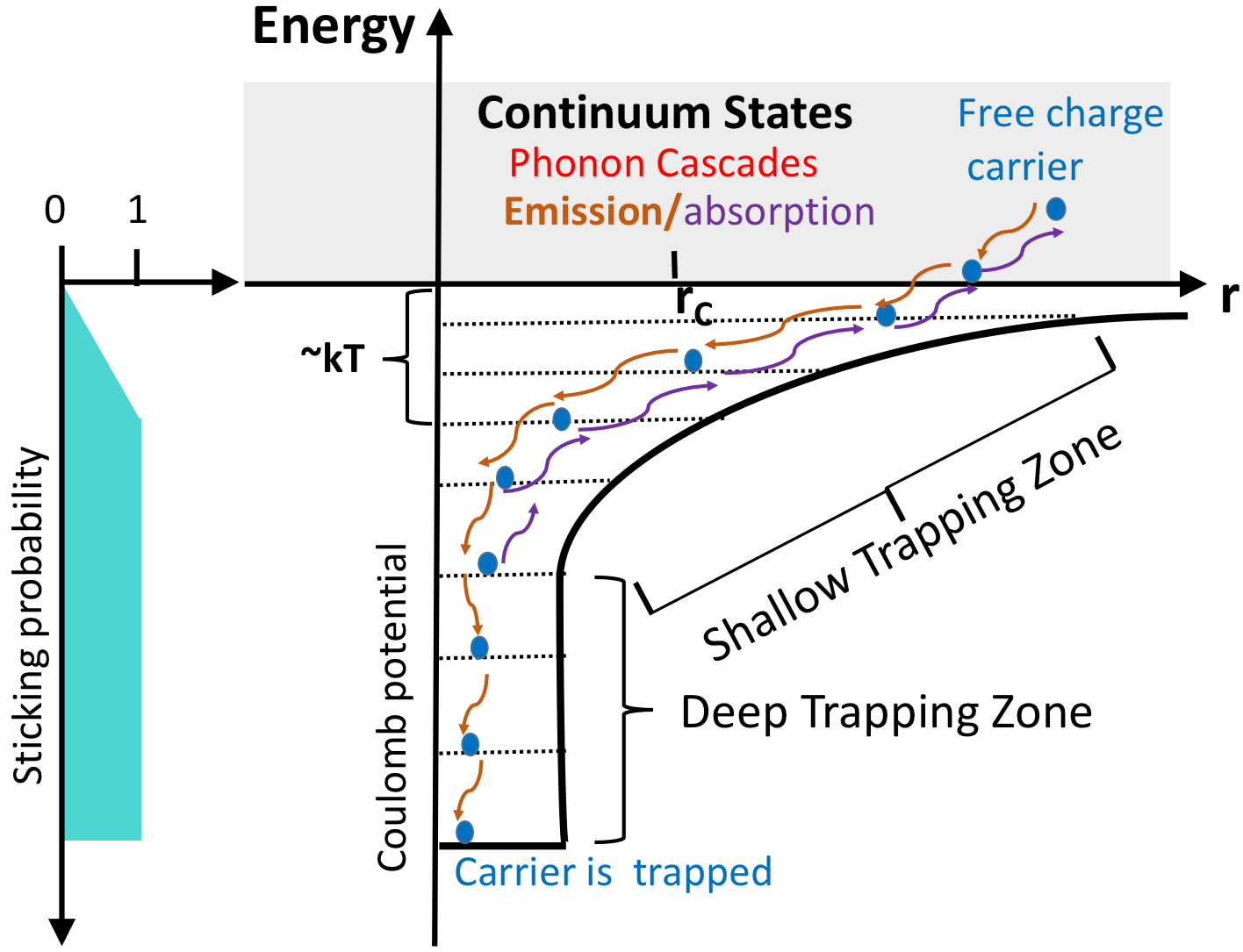}
\caption{\small{Shown is a description of phonon cascade charge trapping processes similar to a plot shown in a thesis from SuperCDMS~\cite{phipps}. $r_{c}$ = $\frac{e^2}{4\pi\epsilon\epsilon_{0}k_{B}T}$ is the so called Onsager radius~\cite{onsager} in which charge carriers can be considered to be bound since their mutual attraction energy $\frac{e^2}{4\pi\epsilon\epsilon_{0}r}>k_{B}T$, where $e$ is the unit of electrical charge, $\epsilon$ is the relative permittivity of Ge,$\epsilon_{0}$ is the permittivity of free space, and $k_{B}$ is the Boltzmann constant.  
 }}
\label{fig:trapping}
\end{figure}

The two trapping processes correspond to two kinds of carrier traps,  namely shallow and deep traps as shown in Figure~\ref{fig:trapping}. This classification is based upon the relative energy levels of the wells defining the traps similar to the low-temperature regime defined by pioneers for CDMS~\cite{sundqvist, phipps}. 

The trapping length is a measure which incorporates both the trapping cross section and the absolute impurity concentration of the detector. For a particular detector thickness, a shorter trapping length is usually indicative of a larger percentage of the charge carriers getting trapped within the detector which would lead to broadening of the resulting peak.
However, the estimation of the trapping length for a given detector is difficult to predict as it depends on the distribution of the traps, which are distributed both spatially and energetically within the semiconductor layers of the detector. This in turn is affected by the electric field distribution within the detector. For complicated detector geometries, accurately estimating the net field distribution and defining a model to calculate the trapping length, and by extension, the absolute impurities is a non-standard process and may involve approximations which contribute to inaccuracies in the results.

With the need to create ever more sensitive detectors which can discriminate both the position and energy of incident $\gamma$ rays simultaneously, various studies have been performed over the years to determine the performance of HPGe detectors which have been fabricated with amorphous germanium (a-Ge) contacts~\cite{21,22,23,24,25,26,27,28}. It has been shown that they are a viable alternative to the traditional and relatively thick n$^+$ contacts formed by Li diffusion for hole blocking and p$^+$ contacts formed by ion hole implantation for electron blocking~\cite{26,27,28}. Various HPGe planar detectors with a-Ge contacts have been fabricated and characterized in-house~\cite{29,30} at the University of South Dakota (USD) with the goal of ultimately being able to build detectors underground in order to avoid cosmogenic activation of any radioisotopes within the detector material~\cite{31} which may contribute to background events during operation.

Planar detectors offer a unique opportunity to study a specific charge trapping process - charge carrier capture which results in the broadening of energy resolution due to the permanent charge loss. This is because the electric field can be precisely calculated for a planar geometry. In addition,  the assumption of an uniform distribution of the impurity is valid with small planar detectors.  Thus, the drift velocity and the total velocity can then be well determined for charge carriers. With the well-determined net impurity level using the depletion length and the depletion voltage through the I-V and C-V curves~\cite{wei}, one can extract the absolute impurity levels for both p-type and n-type trapping centers utilizing the well-established theoretical prediction of the capture cross sections~\cite{lax}. This would allow us to evaluate the impact of charge trapping on the energy resolution for Ge detectors in terms of the absolute impurity levels, which is usually not accessible. 

In this paper, the experimental setup for studying charge trapping in presented in section~\ref{sec:setup} and followed by the description of the charge trapping cross section in sections~\ref{sec:model0}. A model that correlates the energy loss due to charge trapping (charge collection efficiency and charge trapping length) with the measured energy resolution is discussed in section~\ref{sec:model}. The relationship between the charge trapping length and the trapping cross-section (the Lax model) as well ass the absolute impurity is stated in section~\ref{sec:model2} while the results and conclusions are discussed in sections ~\ref{sec:result} and ~\ref{sec:conl}.

\section{Experimental setup}
\label{sec:setup}
\begin{figure} [htbp]
  \centering
  \includegraphics[angle=0,width=14.cm] {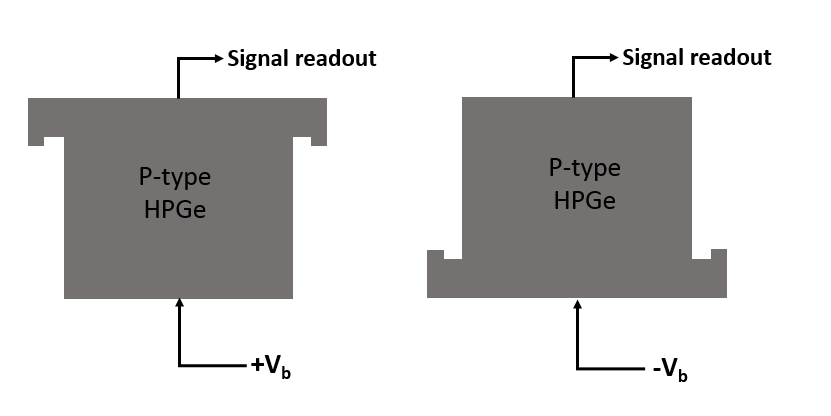}
  \caption{\small{Two orientations for measuring energy resolution of our detectors. Left: the detector is upside down when loaded onto the cryostat, and a positive bias voltage is applied. The holes are collected in this case;  Right: the detector is right side up when loaded onto the cryostat, and a negative bias voltage is applied. The electrons are collected in this case.}}
  \label{fig:setup}
\end{figure}

There are two orientations used for measuring the energy resolution in terms of FWHM in this work, as shown in Figure~\ref{fig:setup}. Both electrons and holes were collected using these two setups, respectively. When a negative bias voltage is applied to the bottom of the detector, the detector is depleted from the top and electrons are drifted across the detector to the side where signal is collected. On the other hand, when the detector is placed upside down, a positive voltage is applied to the top of the detector, the detector is still depleted from the top and holes are drifted to the side where the signal is collected. With these two different orientations, electron trapping and hole trapping are allowed to studied respectively. Table~\ref{tab:t1} summarizes the operational parameters.

\begin{table}[h]
\centering
\caption{A summary of the operational parameters for the nine detectors used in this analysis. $L$ stands for the detector thickness. $V_{b}$ is the applied bias voltage. $V_{d}$ is the depletion voltage. The quoted uncertainty on the net impurity ($|N_A-N_D|$) is mainly due to the error in the determination of the depletion voltage using the I-V and C-V curves.}
\label{tab:t1}
\begin{tabular}{|c|c|c|c|c|}
\hline \hline
Detector & L (cm) & $V_{b}$ (Volts)& $V_d$ (Volts)& $|N_A-N_D|$ (10$^{10}$ cm$^{-3}$) \\ \hline\hline
 USD-W04&1.08&-1500&-300 &0.46$\pm$0.06\\ \hline
 USD-W03&0.94&+2500&+1100 & 2.3$\pm$0.20\\ \hline
 USD-RL01 &1.08&-1500&-400 & 0.62$\pm$0.08\\ \hline
 USD-R02&0.65&-1700&-700 &2.97$\pm$0.04\\ \hline
 USD-R03 &0.81&-2500&-1400 &3.83$\pm$0.07\\ \hline
 USD-L01&0.54&+1500&+650 & 4.00$\pm$0.05\\ \hline
 USD-L06&0.85&+2500&+1200 & 2.98$\pm$0.03\\ \hline
 USD-L07&0.85&+3000&+1000 &2.48$\pm$0.02 \\ \hline
 USD-L08&0.85&+3000&+800 &1.98$\pm$0.01 \\ \hline \hline
\end{tabular}
\end{table}

When drifting charge carriers across the detector, they will meet impurity atoms and become trapped if the mutual attractive potential is sufficient as described in Figure~\ref{fig:trapping}. In the case of deep traps, charge carrier capture occurs as $\e^{-} + D^{+}\longrightarrow D^{0}$ and $h^{+} + A^{-} \longrightarrow A^{0}$, where $D^{+}$ and $A^{-}$ represent the respective spatially localized ionized p-type and n-type impurities in the depleted detectors. $D^{0}$ and $A^{0}$ stand for the neutral states of p-type and n-type impurities, respectively. The trapping mechanism is depicted in Figure~\ref{fig:trapping} and the capture cross section is dependent on the applied electric field, as described below. 

\section{Charge carrier capture cross section}
\label{sec:model0}
In early 1960s, Melvin Lax proposed a phonon cascade mechanism that caused the charge carrier capture in Ge. In his theory, when a charge carrier is drifted to approach an impurity site, it may recombine with this impurity site by the emission of a single, relatively small energy phonon. Due to the Coulomb attraction between the impurity site and the incident charge carrier,  a successive chain of phonons transitioning from a nearly-ionized state to the ground state must occur, as depicted in Figure~\ref{fig:trapping}, to make this carrier become ultimately bound and eventually captured by the impurity center. The distance between the incident carrier and the impurity site for this bound state to occur is the Onsager radius, $r_{c}$ = 13.56 nm for Ge at 77 K. 

On the other hand, as described in Figure~\ref{fig:trapping}, there also exists some probability for phonon absorption to knock the carrier up the energy ladder and back into the continuum states, preventing the carrier from becoming further trapped. This phenomenon is depicted in Figure~\ref{fig:trapping} as the charge carrier falls into a shallow trapping zone. Within this zone, the charge carriers have some probability to be released by absorbing a phonon. These released charge carriers can be corrected for their drift time and hence no longer contribute to the energy broadening of the energy resolution. 

However, as the carrier falls deeper into the Coulomb potential well, the probability for the carrier to become fully captured is high. Melvin Lax introduced an energy-dependent sticking probability which gives the likelihood for the carrier to escape a given bound state~\cite{lax}, as illustrated in Figure~\ref{fig:trapping}. This sticking probability depicts a thermal equilibrium case where an incident charge capture cross section is depicted by a critical radius, $R_{c}$, and is determined by an energy equality, $\frac{e^2}{4\pi\epsilon\epsilon_{0}R_{c}}$=$\frac{3k_{B}T}{2}$. Within a sphere formed by this radius, $R_{c}$, the charge carriers with kinetic energies greater than the thermal equilibrium
average of 3/2 $k_{B}$T usually lose energy through subsequent scattering, whereas
charge carriers with energies much lower than this will gain energy on average. Therefore, the volume determined by this critical radius represents an effective sphere where charge carriers 
are recombined. Given a mean free scattering path, $\lambda(E)_{c}$, the cross section proposed by Melvin Lax is:
\begin{equation}
\label{crosssection}
\sigma_{eff}(E) = \frac{4\pi}{3}\frac{R_{c}^3}{\lambda_{c}(E)},
\end{equation}
where $\sigma_{eff}$ represents the effective cross section. The mean free scattering length, $\lambda_{c}(E)$ = $v_{d}\times\tau_{c}$, where $v_{d}$ is the drift velocity of charge carriers under a given electric field ($E$) and $\tau_{c}$ is the average scattering time within the effective sphere. 

It is well known that the drift velocity, $v_{d}$ within the detector is dependent on the applied electric field~\cite{47,48}. At low fields, there is a linear correlation between the drift velocity and the electric field. While at high fields, the drift velocity varies very slowly with increasing electric fields and reaches saturation beyond a certain point. This phenomenon can be described as: $v_{d}$ = $\mu(E)E$,  
where $\mu(E)$ is defined as the field dependent mobility.
This can be easily incorporated into the relation by using a simple empirical model as below:
\begin{equation}
    v_d \approx \frac{\mu_0E}{1 + \cfrac{E}{E_{sat}}},
    \label{satur}
\end{equation}
where $\mu_0$ is the mobility of the charge carrier at zero field~\cite{49}; 
$E_{sat} = \cfrac{v_{sat}}{\mu_0}$, where $v_{sat}$ is defined as the saturation drift velocity~\cite{50}. Substituting this in equation~\ref{satur}, we obtain the drift velocity as:
\begin{equation}
    v_d = \frac{\mu_0 E}{1 + \cfrac{\mu_0E}{v_{sat}}},
    \label{velsat}
\end{equation}
where $\mu_0$ = $\mu_0(H)/r$ and $\mu_0(H)$ is the Hall mobility. According to the IEEE Standard~\cite{49}, $\mu_0(H)$ = 36000 cm$^{2}/Vs$ and $r$ = 0.83 for electrons, $\mu_0(H)$ = 42000 cm$^{2}/Vs$ and $r$ = 1.03 for holes. The saturation velocity, $v_{sat}$, can be calculated according to an empirical formula below~\cite{50}:
\begin{equation}
    v_{sat} = \frac{v_{sat}^{300}}{1-A_{\nu}+A_{\nu}(T/300)}.
    \label{satvel}
\end{equation}
The parameter values for velocity saturation model~\cite{50} are given as:  $v_{sat}^{300}$ = 0.7$\times$10$^{7}$cm/s for electrons and 0.63$\times$10$^{7}$cm/s for holes for temperature at 300 K, $A_{\nu}$ = 0.55 for electrons and 0.61 for holes. Putting these parameters into equation~\ref{satvel}, $V_{sat}$ = 1.18$\times$10$^{7}$cm/s for electrons and 1.15$\times$10$^{7}$cm/s for holes at 77 K. 

The average scattering time, $\tau_{c}$, can be calculated as~\cite{peter}
\begin{equation}
    \tau_{c} = \frac{m^{*}\mu(E)}{e},
    \label{scatime}
\end{equation}
where $m^{*}$=0.12$m_{0}$ or 0.21 $m_{0}$, is the effective conduction mass of electrons or holes, respectively; $m_{0}$ is the mass of electrons in vacuum; $\mu(E)$ = $\mu_0$/(1+$\mu_0$E/$v_{sat})$. 

\begin{figure}[htb!!!]
\includegraphics[angle=0,width=14.cm] {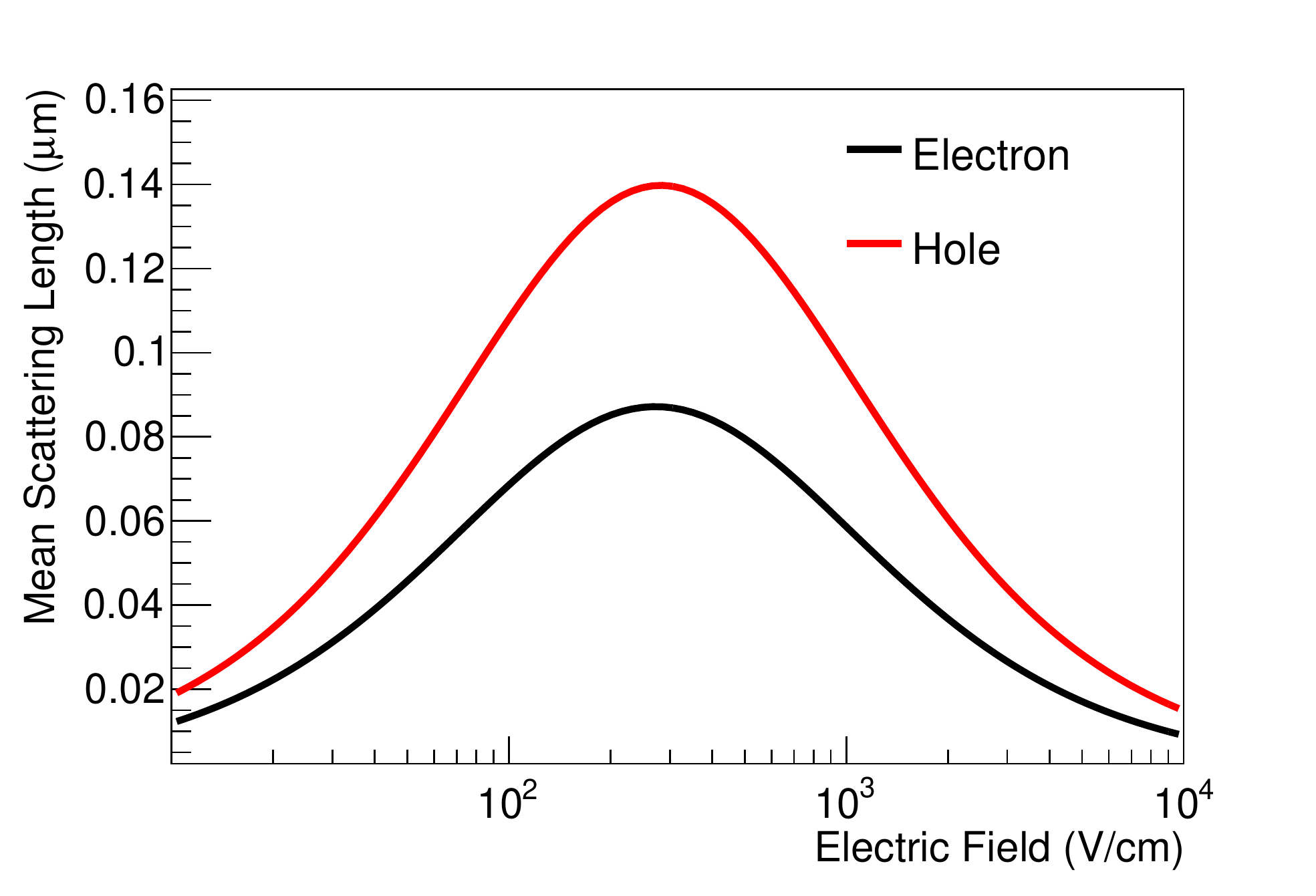}
\caption{\small{Shown is the mean free scattering lengths, $\lambda_{c}(E)$ = $v_{d}\times\tau_{c}$, for electrons and holes as a function of electric field. 
 }}
\label{fig:meanfreepath}
\end{figure}

\begin{figure}[htb!!!]
\includegraphics[angle=0,width=14.cm] {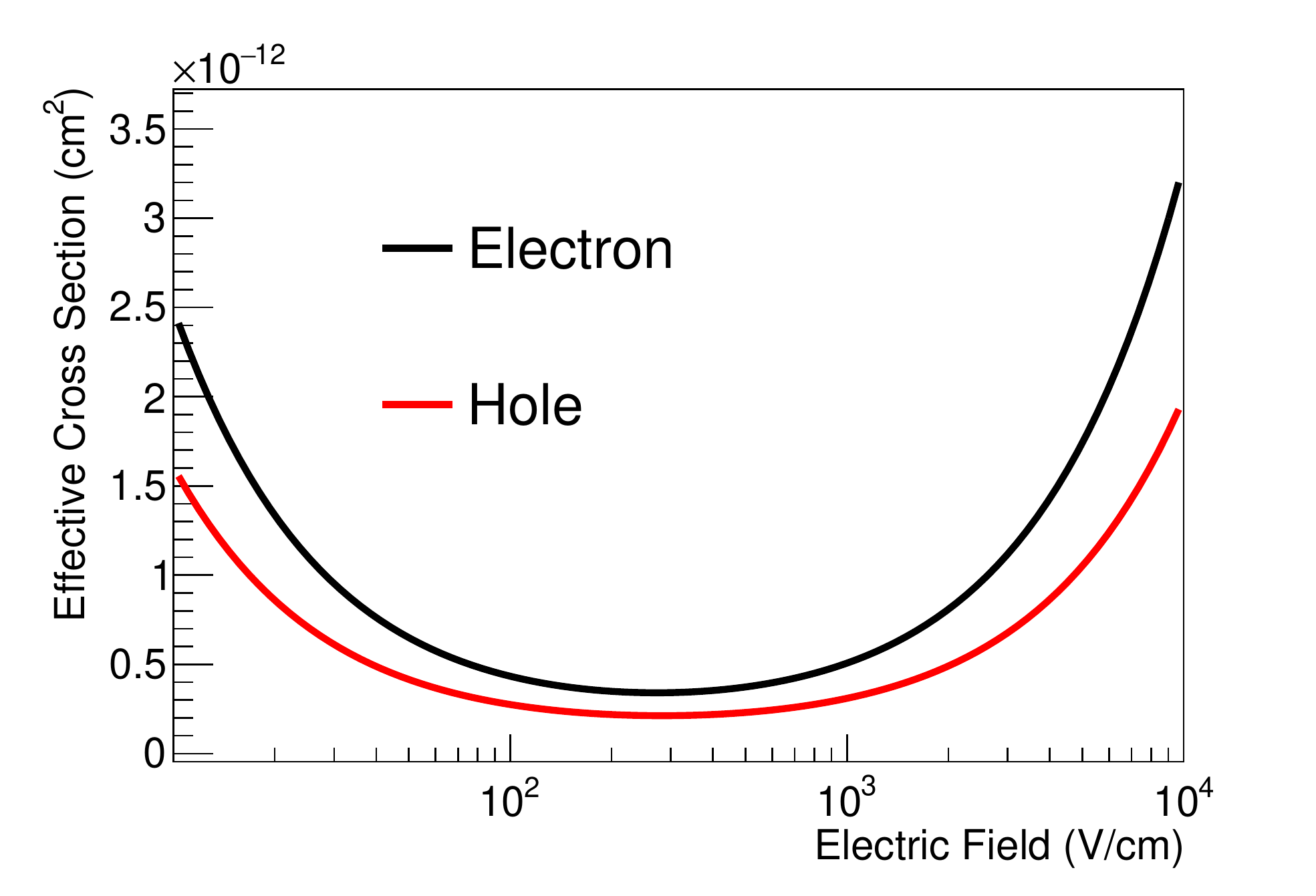}
\caption{\small{Shown is the effective capture cross sections for electrons and holes as a function of electric field. 
 }}
\label{fig:capture}
\end{figure}

Therefore, the field dependent effective capture cross section can be calculated through equations~\ref{crosssection}, ~\ref{velsat}, ~\ref{satvel}, and ~\ref{scatime}. Figure~\ref{fig:meanfreepath} shows the mean free scattering length as a function of the applied electric field for electron and holes. Correspondingly, the effective capture cross sections for electrons and holes can be obtained, as shown in Figure~\ref{fig:capture}. 

Figure~\ref{fig:capture} depicts that electrons have a larger capture cross section than holes. This is because the mean free scattering path of electrons is smaller than that of holes, as shown in Figure~\ref{fig:meanfreepath}. Therefore, it is expected that charge trapping for electrons is more severe than that of holes. Figure~\ref{fig:capture} also implies that applying a higher electric field can enhance deep level charge trapping. Since the loss of charge carriers due to charge trapping is proportional to the number density of trapping centers, the observation of charge trapping would allow us to find out the absolute impurity level of Ge crystals, which were used to fabricate Ge detectors. This provides a vital value in evaluating Ge detectors for rare-event physics searches. 

A direct consequence of charge trapping is to lose charge collection efficiency during the drifting process. As a result, the total collected charge carriers is less than the total generated charge carriers. This implies that charge trapping will impact the measured energy resolution.

\section{The correlation between charge collection efficiency and the actual energy resolution}
\label{sec:model}
\subsection{Energy resolution}
The energy resolution, $\Delta$E of a HPGe detector is defined as the width of a characteristic $\gamma$ peak which corresponds to half the maximum energy values on either side of the peak. It consists of three components, namely:
$\Delta E_{sv}$ $-$ the energy resolution due to statistical variation alone;
 $\Delta E_{ic}$ $-$ the energy resolution due to incomplete charge collection or charge trapping, as depicted in Figure~\ref{fig:trapping};
$\Delta E_{en}$ $-$ the energy resolution due to electronic noise.
These three components are related to each other through the equation below:
\begin{equation}
\label{e2}
    \Delta E^{2} = \Delta {E_{sv}}^2 + \Delta {E_{ic}}^2 + \Delta {E_{en}}^2.
\end{equation}
Usually, the terms $\Delta E_{sv}$ and $\Delta E_{ic}$ are tangled together while taking observations as $\Delta E_{ic}$ is usually small. For a given planar detector, $\Delta E$ and $\Delta E_{en}$ are measurable quantities and can be obtained with relative ease for a particular energy peak as discussed in section ~\ref{sec:result}. Hence, the  convolution of $\Delta E_{sv}$ and $\Delta E_{ic}$ for the particular energy peak can be given by: 
\begin{equation}
\label{e3}
    \Delta E_{m} = \sqrt{\Delta E^2 - \Delta {E_{en}}^2},
\end{equation}
where $\Delta E_{m}$ = $\sqrt{\Delta E_{sv}^{2} + \Delta E_{ic}^2}$ depends on the Fano factor, $F$ (which is material specific), the energy loss due to deep level trapping (as depicted in Figure~\ref{fig:trapping}.), the average energy, $\epsilon$, needed for the production of one electron-hole (e-h) pair at the given operating temperature and the energy, $E$, of the absorbed $\gamma$ ray, and the impact ionization of impurities if detectors  are operated at high field. 

\subsection{Charge collection efficiency}
In order to define a suitable model which relates the above parameters to allow one to understand the fraction of energy loss due to deep level trapping, we make use of the Shockley-Ramo theorem~\cite{43} which relates the instantaneous current, $i$, induced on a given electrode of the planar  detector due to the motion of charges, to electric field and is given by:
\begin{equation}
\label{e30}
    i = {E_v}qv,
\end{equation}
where $q$ denotes the charge of the particle; $v$ represents the instantaneous velocity;
$E_v$ stands for the component of the weighting electric field or Ramo field (note that this is not the electric field.) in the direction of $v$ at the charge's instantaneous position.

\begin{figure}[htb!!!]
\includegraphics[angle=0,width=10.cm] {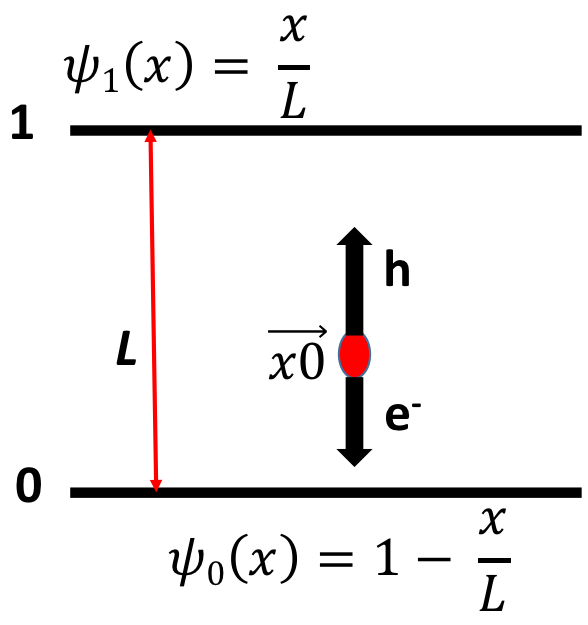}
\caption{\small{Shown is a sketch of the drifting process and the weighting potential for a planar detector geometry. $x$ is the position displacement along the z-axis, the direction of the electric field. 
 }}
\label{fig:planar}
\end{figure}

Figure~\ref{fig:planar} depicts the drifting process and the weighting potential for a planar geometry. The bottom of the detector is biased with a voltage which is much higher than the full depletion voltage necessary. This ensures that the active volume of the detector is equal to its thickness of $L$.

As given by equation~\ref{e30}, we know
\begin{equation}
\label{e4}
    i = {E_v}qv = {E_v}q\frac{dx}{dt}.
\end{equation}
Consider the bottom plate wherein the electric field, $E_v$ is equal to:
\begin{equation}
\label{e5}
    E_v = -\nabla {\psi_0(x)} = \frac{1}{L}.
\end{equation}
From the basic definition of current, we know that 
\begin{equation}
\label{e6}
    i = \frac{dQ}{dt}, i.e., dQ = idt.
\end{equation}
Substituting equations~\ref{e4} and \ref{e5} into \ref{e6},  we obtain
\begin{equation}
\label{e7}
 dQ = \frac{dx}{L}q.   
\end{equation}
 For an electron, $q = -q$; Hence, the equation can be re-written as:
\begin{equation}
\label{e8}
 dQ = -\frac{dx}{L}q.    
\end{equation} 

 Let us assume a population of $N_h(0)$ holes at one end of the electrode (x = 0) travelling towards the other end which is a distance $x = L$ apart.
 Due to the presence of impurities within the Ge crystal, some of these charges will be trapped as they move upwards. The trapping mechanism is described in Figure~\ref{fig:trapping}. Keep in mind that only the deep trapping mechanism is relevant to a planar detector when the over-biased voltage is applied, which guarantees a highly field sufficient to drift charge carriers across the detector.

 The mean length along the z-axis that charge carriers travel through before they are trapped is defined as trapping length and is denoted by $\lambda_{th}$. Trapping length is detector specific. Hence the population of the holes as they traverse the detector will fall exponentially in accordance with the exponential distribution that is used to characterize the dispersion of traps in the forbidden energy gap as proposed by Rose~\cite{44}. The population of holes, $N_h(x)$, at a distance $\Vec{x} = x$ is given as:
 \begin{equation}
 \label{e9}
     N_h(x) = N_h(0)e^{-\frac{x}{\lambda_{th}}}.
 \end{equation}
 
 The induced charge at the electrode where the charge is collected for any position, $x$, within the planar detector is thus:
 \begin{equation}
 \label{e10}
     Q_{h,x} = \int^L_0{qN_h(x)\frac{dx}{L}}.
 \end{equation}
 Integrating the drift path from $x=0$ to $x=L$, we obtain:
 \begin{equation}
    Q_{h,L} =  qN_h(0)\frac{\lambda_{th}}{L}(1 - e^{-\frac{L}{\lambda_{th}}}).
 \end{equation}
 Hence,
  \begin{equation}
  \label{11}
     Q_{h,L} =  Q_{h,0}\frac{\lambda_{th}}{L}(1 - e^{-\frac{L}{\lambda_{th}}}),
 \end{equation}
 where $Q_{h,L}$ = total charge at x = L; $Q_{h,0}$ = total charge at x = 0;
 It is evident from the above equation that the total charge at $\Vec{x} = L$ is less than that at $x = 0$. Thus, the ratio of the total charge at x= 0, $Q_{h,0}$, to the total charge at x=$L$, $Q_{h,L}$, is defined as the charge collection efficiency for a given planar detector:
 \begin{equation}
     \label{e100}
     \varepsilon_{h} = \frac{Q_{h,L}}{Q_{h,0}} = \frac{\lambda_{th}}{L}(1 - e^{-\frac{L}{\lambda_{th}}}).
 \end{equation}
 A similar equation can be derived for electrons. Therefore, equation~\ref{e100} can be used to study the charge collection efficiency for both electrons and holes. 
 
 \subsection{The relation between energy resolution and charge collection efficiency}
 Without charge trapping, the energy resolution after the subtraction of electronic noise, $\Delta E_{m}$ = $\Delta E_{sv}$, is given by:
 \begin{equation}
 \label{12}
     \Delta E_{sv} = 2.355\sqrt{FE\epsilon}
= \frac{2.355E\sqrt{F}}{\sqrt{\cfrac{E}{\epsilon}}},
\end{equation}
where $F$ is the Fano factor, $E$ is the total energy of deposition in the detector, $\epsilon$ is the average energy required to produce one e-h pair. If one assumes $N_{tot}$ = $\frac{E}{\epsilon}$, therefore, equation~\ref{12} becomes:
\begin{equation}
\label{13}
\Delta E_{sv} = \frac{2.355E\sqrt{F}}{\sqrt{N_{tot}}},
\end{equation}
  where $N_{tot}$ is the total number of electron-hole pairs generated.

However, because of deep level trapping, the total number of charge carriers that are actually collected at the opposite plate is: $N_{tot} = \cfrac{E}{\epsilon}\varepsilon_h$. Therefore, 
the energy resolution after the subtraction of electronic noise, $\Delta E_{m}$ = $\sqrt{\Delta E_{sv}^{2} + \Delta E_{ic}^{2}}$, is given by:
\begin{equation}
\label{ee14}
    \Delta E_{m} = 2.355\sqrt{EF\epsilon/\varepsilon_{h}}.
\end{equation}
To accurately study the relation between the energy resolution of statistical variation and charge trapping when running the detectors at high electric fields, one must take into account the impact ionization of impurities. Due to a relative low-energy ionization threshold ($\sim$0.01 eV) of impurities in Ge, hot charge carriers can ionize impurity atoms to gain more charge carriers. Thus, a correction factor ($g_c$) to the total number of charge carriers that are actually collected at the opposite plate should be applied, which results in $\frac{N_{tot}}{g_{c}}$ = $\frac{E}{\epsilon}\varepsilon_{h}$. Therefore, equation~\ref{ee14} can be rewritten as:
\begin{equation}
    \label{14}
     \Delta E_{m} = 2.355\sqrt{EF\epsilon/\varepsilon_{h}g_{c}},
\end{equation}
where $g_{c}$ represents a gain factor of charge carriers due to the impact ionization of impurities; $\Delta E_{m}$ and $E$ are directly measurable by analyzing the corresponding $\gamma$-ray spectra; $\epsilon$ $\approx$ 2.96$eV$ for Ge detectors operating at liquid nitrogen temperature. To obtain the charge collection efficiency, $\varepsilon_{h}$, using equation~\ref{14}, one must first determine the Fano factor and the gain factor due to the impact ionization of impurities.  

\subsubsection{Fano factor}
The Fano factor, $F$, is calculated to be $\cong 0.13$ from a theoretical model~\cite{45} for Ge detectors. The value of $F$ has also been reported to be in the range from 0.06 to 0.13 by many experiments~\cite{knoll, ex1, ex2, ex3, ex4}. The discrepancy between those reported values can be explained by equation~\ref{14}, where the charge collection efficiency and the impact ionization were not taken into account in the previous theoretical model~\cite{45} and the experimental determination of the Fano factor using the energy resolution of statistical variation~\cite{knoll, ex1, ex2, ex3, ex4}.

The Fano factor describes the variation of energy dissipation in a collision between ionization and excitation. Therefore, the Fano factor is only related to the creation of charge carriers. Upon deposition of energy in a given target, $E_{0}$, two types of excitations, (a) lattice excitations with no
formation of mobile charge pairs and (b) ionizations with formation of mobile charge pairs, are created by an incoming particle.  
Lattice excitations produce $N_{x}$ phonons of energy $E_{x}$. Ionizations form $N_{i}$ charge pairs of energy $E_{i}$.
For an energy loss process, energy conservation requires $E_{0}$ = $E_{i}N_{i}$ + $E_{x}N_{x}$. As fluctuations in 
$N_{i}$ are compensated by fluctuations in $N_{x}$ to keep $E_{0}$ constant, 
$\frac{dE_{0}}{dN_{i}}\Delta N_{i}$ + $\frac{dE_{0}}{dN_{x}}\Delta N_{x}$ = 0 and $E_{i} \Delta N_{i}$ + $E_{x}\Delta N_{x}$ = 0. 
From averaging many events, 
one obtains for the variance: $E_{i}\sigma_{i}$ = $E_{x}\sigma_{x}$, with $\sigma_{x}$ = $\sqrt{N_{x}}$ assuming Gaussian statistics. 
Thus, $\sigma_{i}$ = $\frac{E_{x}}{E_{i}}\sqrt{N_{x}}$. Since $N_{x}$ = $\frac{E_{0}-E_{i}N_{i}}{E_{x}}$, one obtains
$\sigma_{i}$ = $\frac{E_{x}}{E_{i}}\sqrt{\frac{E_{0}}{E_{x}}-\frac{E_{i}}{E_{x}}N_{i}}$. Each ionization leads to a charge pair 
that contributes to the signal, therefore, $N_{i}$ = $\frac{E_{0}}{\epsilon_{i}}$ and 
\begin{equation}
\sigma_{i} = \sqrt{\frac{E_{0}}{\epsilon_{i}}}\cdot\sqrt{\frac{E_{x}}{E_{i}}(\frac{\epsilon_{i}}{E_{i}}-1)},
\label{eq:sigma}
\end{equation}
 where $\epsilon_{i}$ represents
mean energy expended per e-h pair. 

The statistical variation is usually quantified by the Fano factor~\cite{fano} ($F$), which is defined for any integer-valued 
random variable as the 
ratio of the variance ($\sigma_{i}^{2}$) of the variable to its mean ($N_{i}$), $F$ = $\frac{\sigma_{i}^{2}}{N_{i}}$.  
Thus, one obtains $\sigma_{i}$ = $\sqrt{FN_{i}}$. Comparing this expression to equation~\ref{eq:sigma}, one finds 
\begin{equation}
F = \sqrt{\frac{E_{x}}{E_{i}}(\frac{\epsilon_{i}}{E_{i}}-1)}, 
\label{fano}   
\end{equation}
where $E_{x}$ = 0.0027 eV is the exciton binding energy in Ge~\cite{ggm}; $E_{i}$ = 0.73 eV is the bandgap energy of Ge; and $\epsilon$ = 2.96 eV. Putting all of these numbers into equation~\ref{fano}, we find $F$ = 0.106 for Ge. 

\subsubsection{The impact ionization of impurities} 
As we stated in an earlier publication~\cite{mei}, the ionization energies for the most abundant impurities such as aluminum, gallium, boron, and phosphorus in Ge are in a range of $\sim$0.01 eV. With high electric fields, it is possible that a charge carrier can gain sufficient kinetic energy while drifting across the detector to generate more charge carries~\cite{nsc, jfp}. The gain factor, $g_{c}$, which describes the fraction of charge carriers gained due to the impact ionization of impurities, can be obtained through equation below~\cite{wmo}:
\begin{equation}
    \label{gain}
    g_{c} = 1.0 + \frac{e\lambda_{R}E}{\lambda_{c}E_{x}}exp(-\frac{E_{i}}{e\lambda_{R}E})\times d,
\end{equation}
where $\lambda_{R}$ = $v_{s}\times\tau_{ph}$ is the mean free scattering length between phonons and charge carriers; $v_{s}$ = 5.4$\times$10$^{5}$cm/s is the speed of phonons in Ge and $\tau_{ph}$ =$\mu(E)\times$m$^{*}/e$ is the mean scattering time between phonons and charge carriers; $\lambda_{c}$ is defined in equation~\ref{crosssection}; $E$ is the electric field; $E_{x}$ = 0.0027 eV is described in equation~\ref{fano}; $E_{i}$ = 0.01 eV is the ionization energy of impurities; and $d$ is the actual drifting distance of charge carriers inside the detector. 

Once the charge collection efficiency is determined from the measured energy resolution, one can calculate the average charge trapping length using equation~\ref{e100}. It is commonly known that charge trapping length is related to the effective charge capture cross section and the density of the charge trapping centers. This is further discussed below. 

\section{The relation between charge trapping length, the effective capture cross section, and the absolute impurity level}
\label{sec:model2}
In the case of a uniform distribution of impurities with a number density $N$ = $N_A+N_D$, the charge carrier can become locally captured onto one of these impurities, where $N_A$ is the number density of p-type impurities and $N_D$ is the number density of n-type impurities. Consider the energy-dependent average cross section for this process at a fixed electric field strength by $\bar{\sigma}_{eff}(E)$. Hence, using the measured energy resolution and equation~\ref{14}, we can obtain the charge collection efficiency if the Fano factor and the gain factor are given. Subsequently, utilizing equation~\ref{e100}, we can determine the average trapping length for a given detector. Once the trapping length, $\bar{\lambda}_{th}$, is determined, the average effective trapping cross-section (equation~\ref{crosssection}), $\bar{\sigma}_{eff}(E)$, can be related to the absolute impurity through the following relation~\cite{phipps}:
\begin{equation}
\label{trappinglength}
    \bar{\lambda}_{th} = \frac{1}{(N_A+N_{D}\pm|N_{A}-N_{D}|)/2\times\Bar{\sigma}_{eff}(E)\times\frac{<v_{tot}>}{<v_{d}>}},
\end{equation}
where $N_A+N_D$ is the absolute impurity concentration of the detector; $|N_{A}-N_{D}|$ is the net impurity; $<v_{tot}>$ is the expectation value of the total velocity over the carrier energy distribution; $<v_{d}>$ is the expectation value of the drift velocity over the carrier energy distribution; the $``+"$ sign corresponds to hole trapping and the $``-"$ sign corresponds to electron trapping.  

It is often assumed that the velocity distribution of charge carriers can be approximated by a displaced Maxwellian. Thus the velocity $\textbf{v}$ = $\bf{v_d} + \bf{v_{th}}$,
where $\bf{v_{th}}$ is the thermal velocity with a random direction. The total velocity is then defined as $v_{tot}$ = $<\sqrt{\bf{v}\cdot\bf{v}}>$. 

The electric field within a planar detector can be calculated precisely as below~\cite{46}:
\begin{equation}
\label{efield}
    E = \frac{V_b}{L} + \frac{e|N_A-N_D|}{\epsilon_0\epsilon_r}(\frac{L}{2} - x),
\end{equation}
where $V_{b}$ is the applied bias voltage,  $\epsilon_r$ = 16.2 is the relative permittivity for Ge; $\epsilon_0$ = $8.854\times10^{-14} F/cm$ is the permittivity of free space; $e$ denotes the charge of an electron; $|N_{A}-N_{D}|$ = 2$\epsilon\epsilon_{0}V_{d}/L^2$; $V_{d}$ is the full depletion voltage determined by C-V measurements; $L$ is the thickness of the depletion region which is equal to the detector thickness when operated in full depletion mode. Note that $V_d$ is determined experimentally by measuring the capacitance versus the applied voltage. This can be understood as following. As the bias voltage of the detector, $V_b$, goes up, the thickness of the depleted region, $d$, increases, the detector capacitance, $C_d$, goes down, because $C_d$ is anti-proportional to $d$. When the detector is fully depleted, $d=L$ (the thickness of the detector) cannot increases any more, $C_d$ becomes a constant thereafter. The bias voltage at the point where the $C_d$ versus $V_b$ curve starts to flatten out is therefore the depletion voltage, $V_d$. The uncertainty on $|N_A-N_D|$ is stated in Table~\ref{tab:t1} for all detectors used in this paper. The level of uncertainty is about 13\% for two detectors (USD-W04 and USD-RL01). The remaining seven detectors have a level of uncertainty on $|N_A-N_D|$ within a few percent. Taking a 13\% uncertainty on $|N_A-N_D|$ to calculate the electric field as a function of the bias voltage, $V_b$, the resulted uncertainty is less than 1\%, which is negligible. The accuracy of measuring $V_b$ is within 1\% according to our calibration. 

With established analysis framework in terms of the charge trapping mechanism, charge capture cross section, charge collection efficiency and its relation with the measured energy resolution, and the charge trapping length and its correlation with the absolute impurity level, one can analyze data with nine detectors. Since charge capture cross section is dependent on the applied electric field, hence the charge collection efficiency and the measured energy resolution are also coupled to the applied electric field, the data analysis is proceeded with the calculation of the electric field.  We show the results and discussions below. 

\section{Results and discussions}
\label{sec:result}
\subsection{Electric field and velocity distributions}
\begin{figure}[htb!!!]
\includegraphics[angle=0,width=14.cm] {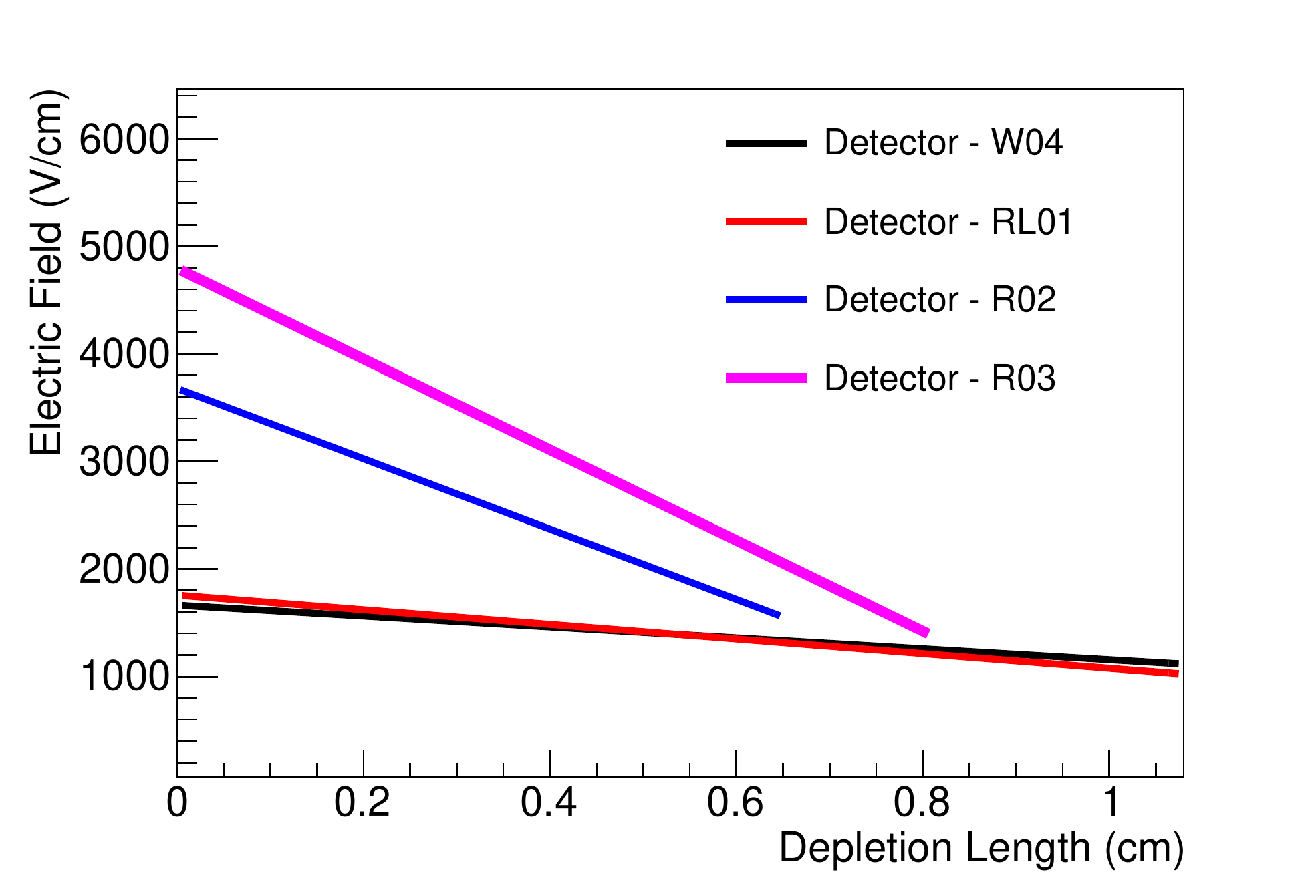}
\caption{\small{Shown is the electric field distribution in the four detectors, which electron trapping is studied. 
 }}
\label{fig:fielde}
\end{figure}

\begin{figure}[htb!!!]
\includegraphics[angle=0,width=14.cm] {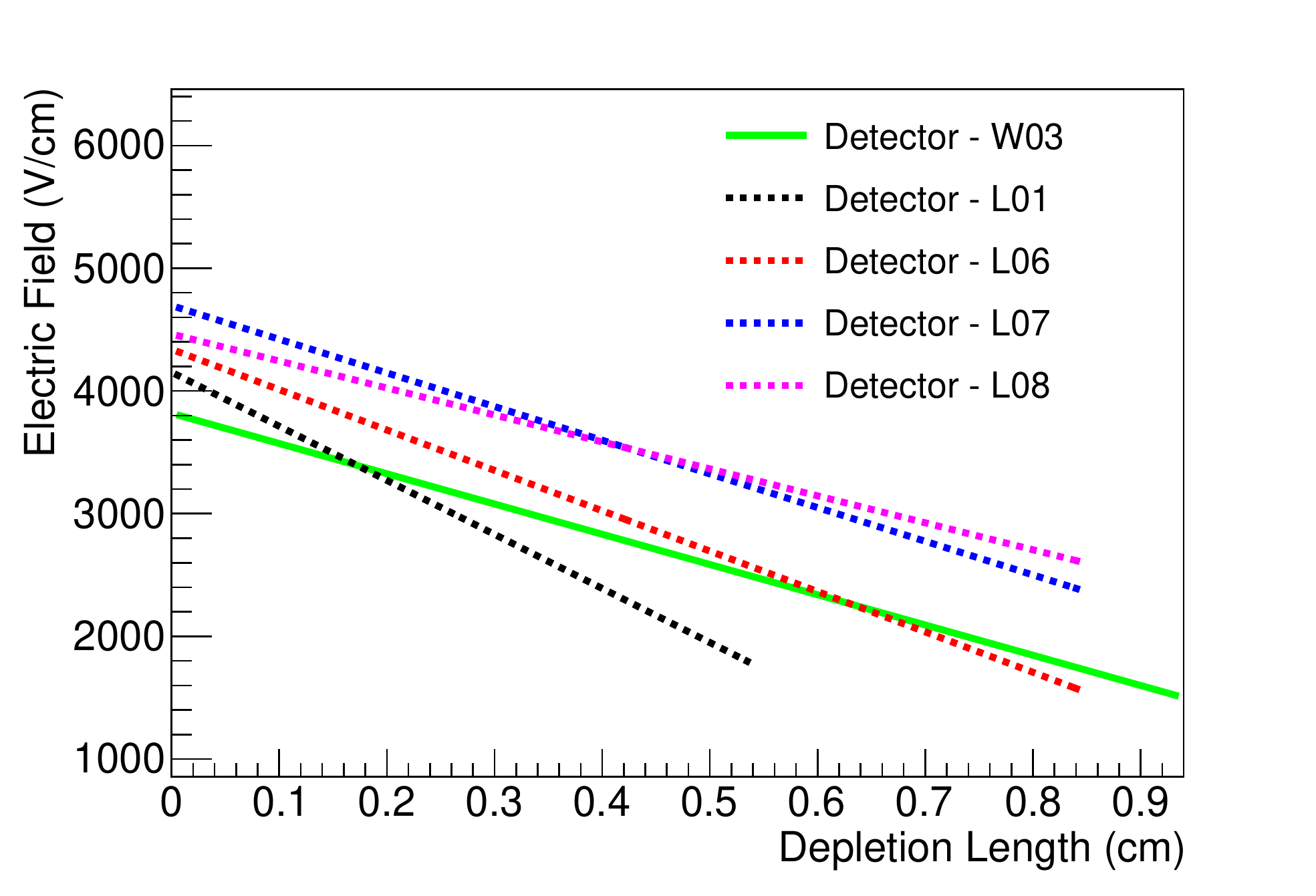}
\caption{\small{Shown is the electric field distribution in the five detectors, which hole trapping is studied. 
 }}
\label{fig:fieldh}
\end{figure}

Nine small planar detectors were used to conduct this study. Utilizing equation~\ref{efield}, we calculate the electric field distribution for each detector, as shown in Figures~\ref{fig:fielde} and \ref{fig:fieldh}.
 Note that a uniform distribution of impurities is assumed when calculating the electric field for each detector. This is quite accurate for small planar detectors at which the gradient of impurities is small. 
 
\begin{figure}[htb!!!]
\includegraphics[angle=0,width=14.cm] {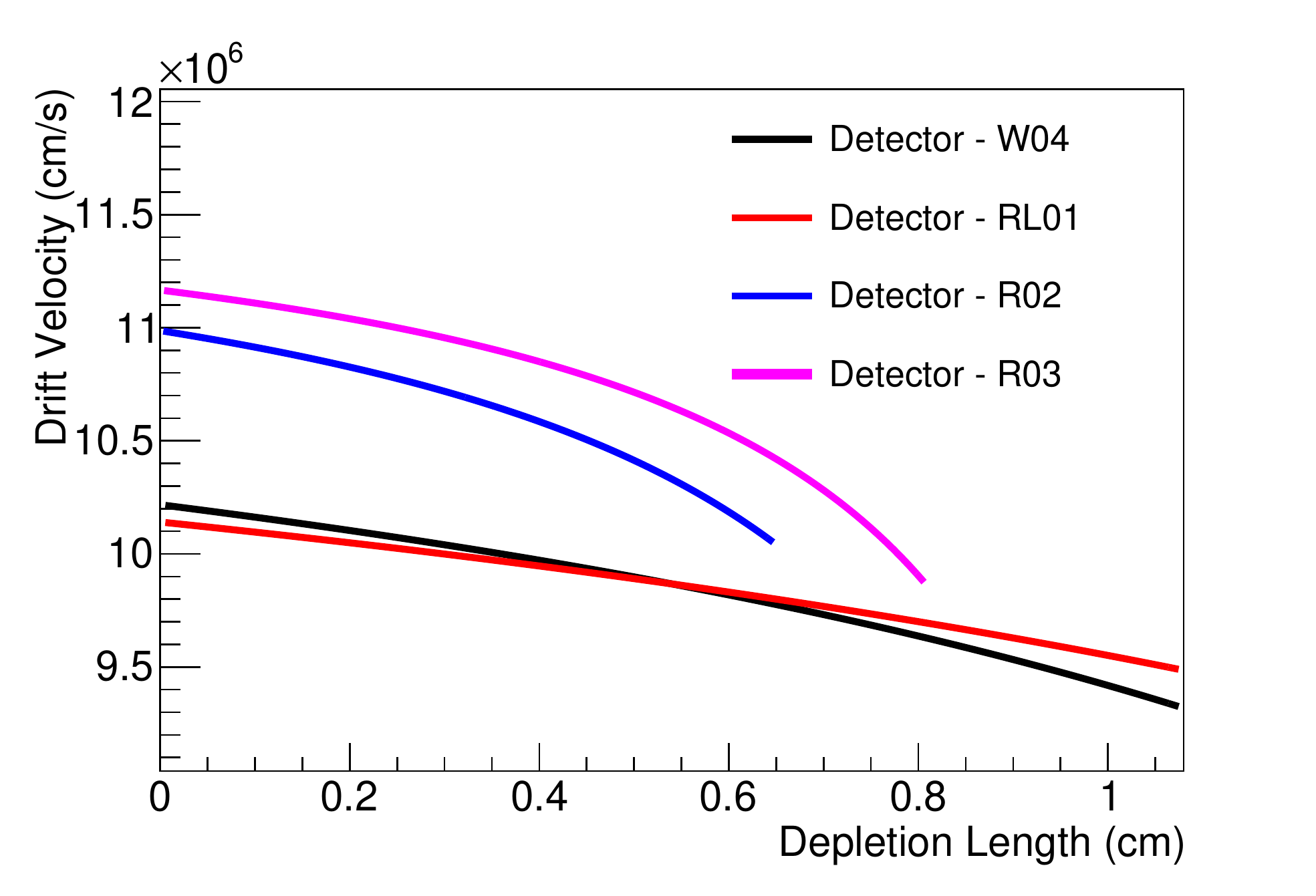}
\caption{\small{Shown is the drift velocity distribution for the four detectors used in studying electron trapping. 
 }}
\label{fig:drifte1}
\end{figure}

\begin{figure}[htb!!!]
\includegraphics[angle=0,width=14.cm] {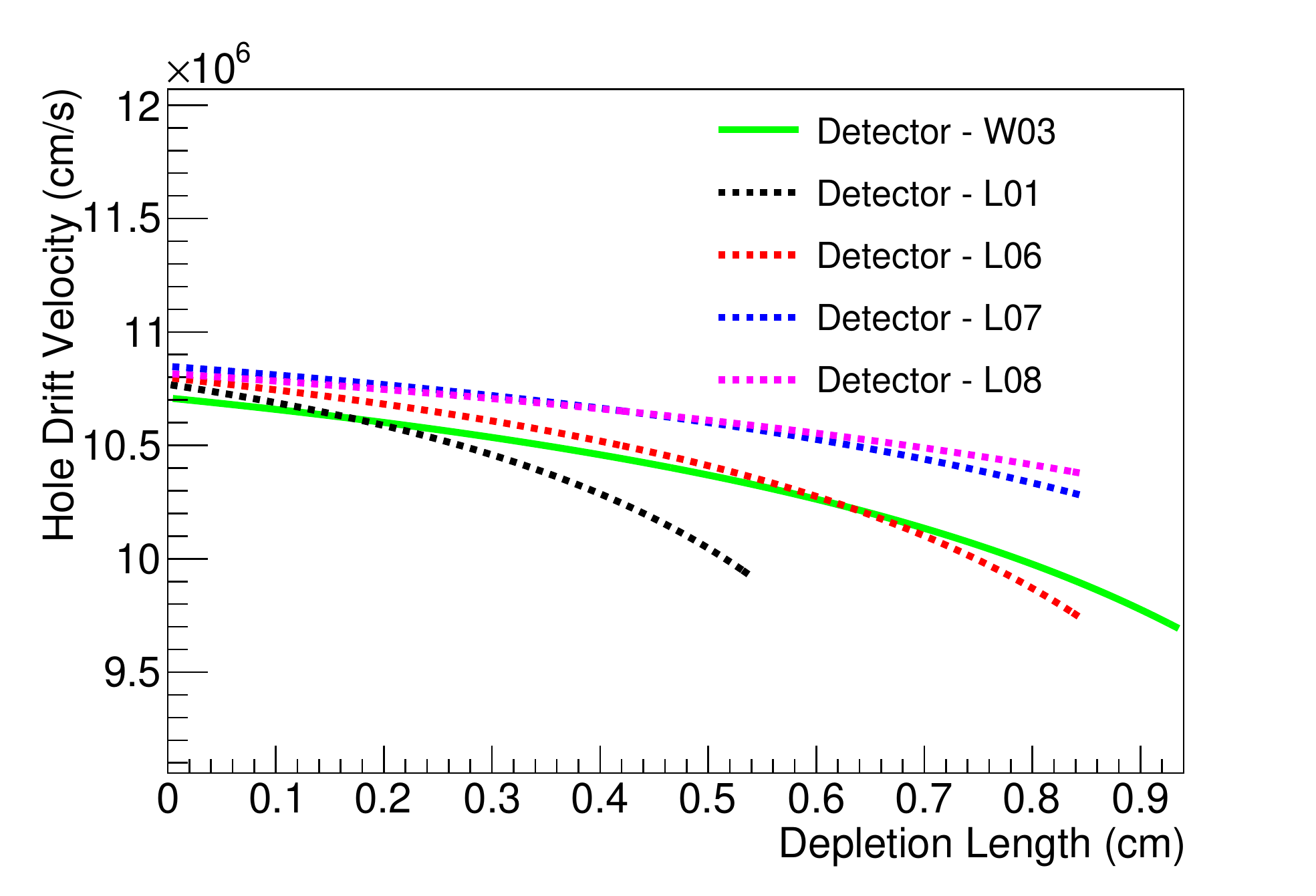}
\caption{\small{Shown is the drift velocity distribution for the five detectors used in studying hole trapping. 
 }}
\label{fig:drifth1}
\end{figure}

\begin{figure}[htb!!!]
\includegraphics[angle=0,width=14.cm] {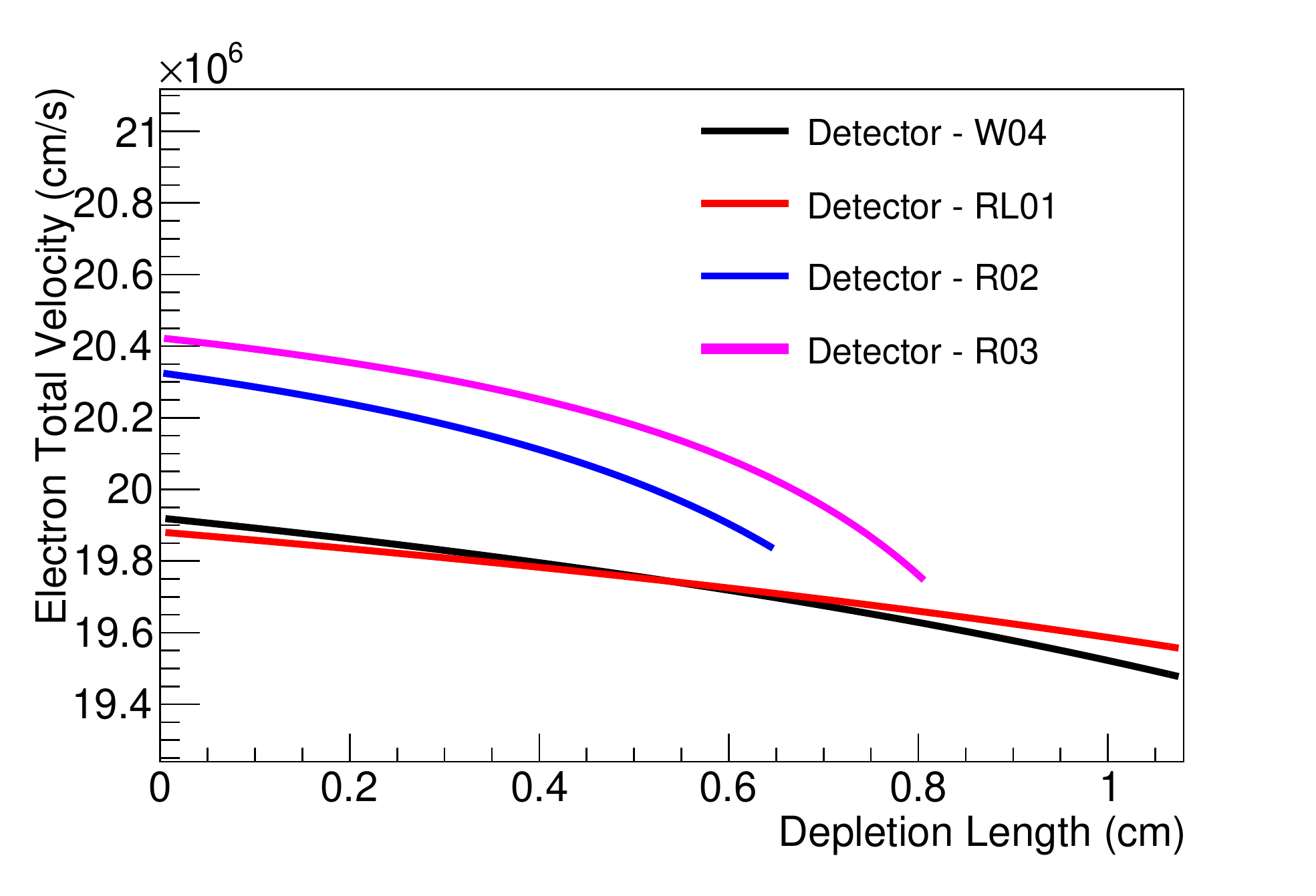}
\caption{\small{Shown is the total velocity distribution for the four detectors used in studying electron trapping. 
 }}
\label{fig:totale1}
\end{figure}

\begin{figure}[htb!!!]
\includegraphics[angle=0,width=14.cm] {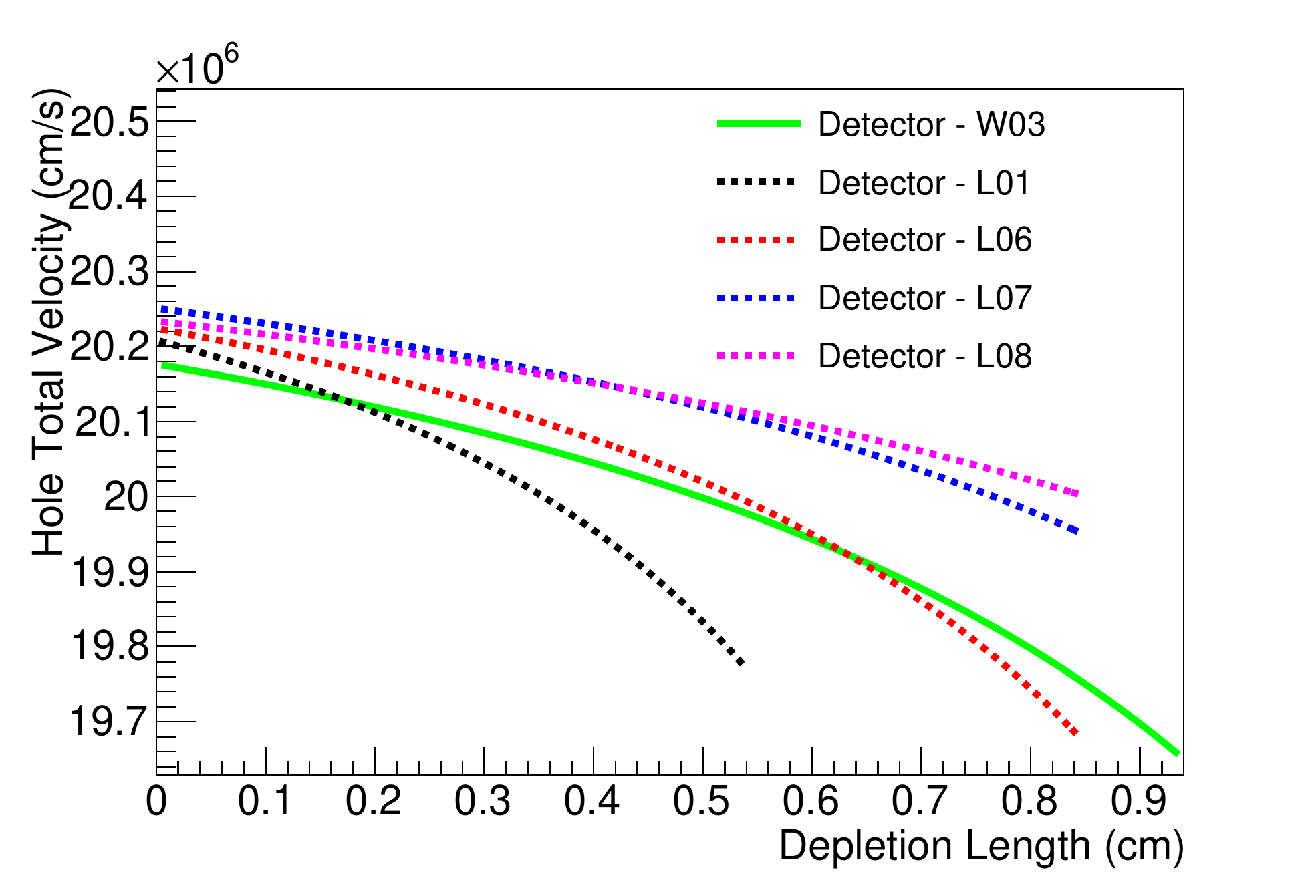}
\caption{\small{Shown is the total  velocity distribution for the five detectors used in studying hole trapping. 
 }}
\label{fig:totalh1}
\end{figure}

It is clear that all nine detectors were operated with sufficient electric field to avoid shallow trapping and increase the deep level trapping probability. 
With a well-understood electric field distribution for each detector, the drift velocity distribution for each detector can be calculated using equation~\ref{velsat}. Hence, the total velocity distribution  can be obtained for each detector with $v_{tot}$ = $<\sqrt{\bf{v}\cdot\bf{v}}>$. Figures~\ref{fig:drifte1}, \ref{fig:drifth1}, \ref{fig:totale1}, and \ref{fig:totalh1} show the distributions for all nine detectors. 
It is noticed that the trends of the drift velocity and the total velocity distributions reflect the electric field distribution inside the detector. The high drift velocity ($>$9.5$\times$10$^{6}$cm/s) in all detectors guarantees a drift time of $\sim$100 ns for a $\sim$1 cm thickness detector. It indicates that the slow pulses due to shallow trapping are largely avoided when the charge collection time is of $\sim$$\mu$s.  

\subsection{Effective mean free scattering path and capture cross sections}
\begin{figure}[htb!!!]
\includegraphics[angle=0,width=14.cm] {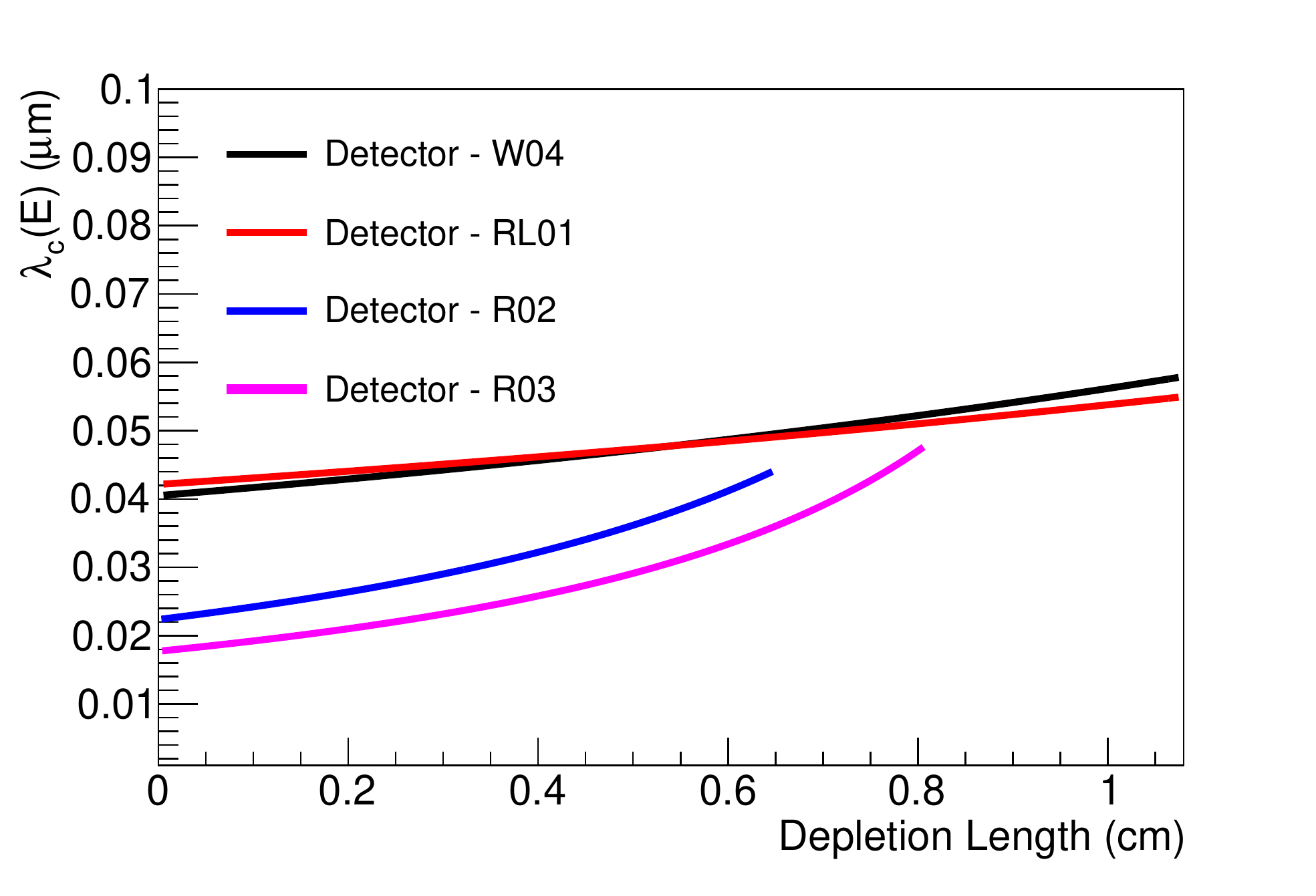}
\caption{\small{Shown is the mean free scattering path distribution for the detectors used in studying electron trapping. 
 }}
\label{fig:meanfe}
\end{figure}

\begin{figure}[htb!!!]
\includegraphics[angle=0,width=14.cm] {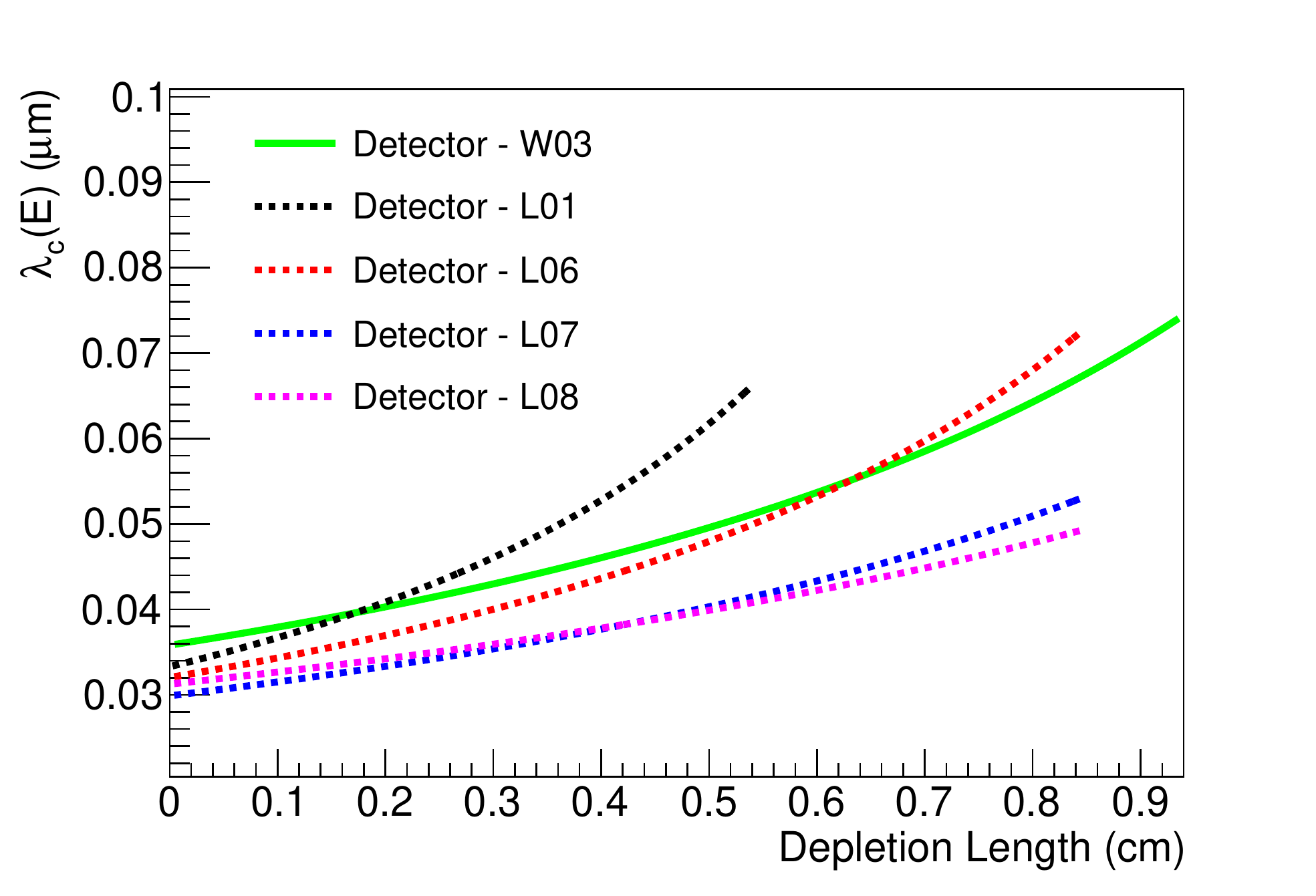}
\caption{\small{Shown is the mean free scattering path distribution for the detectors used in studying hole trapping. 
 }}
\label{fig:meanfh}
\end{figure}

\begin{figure}[htb!!!]
\includegraphics[angle=0,width=14.cm] {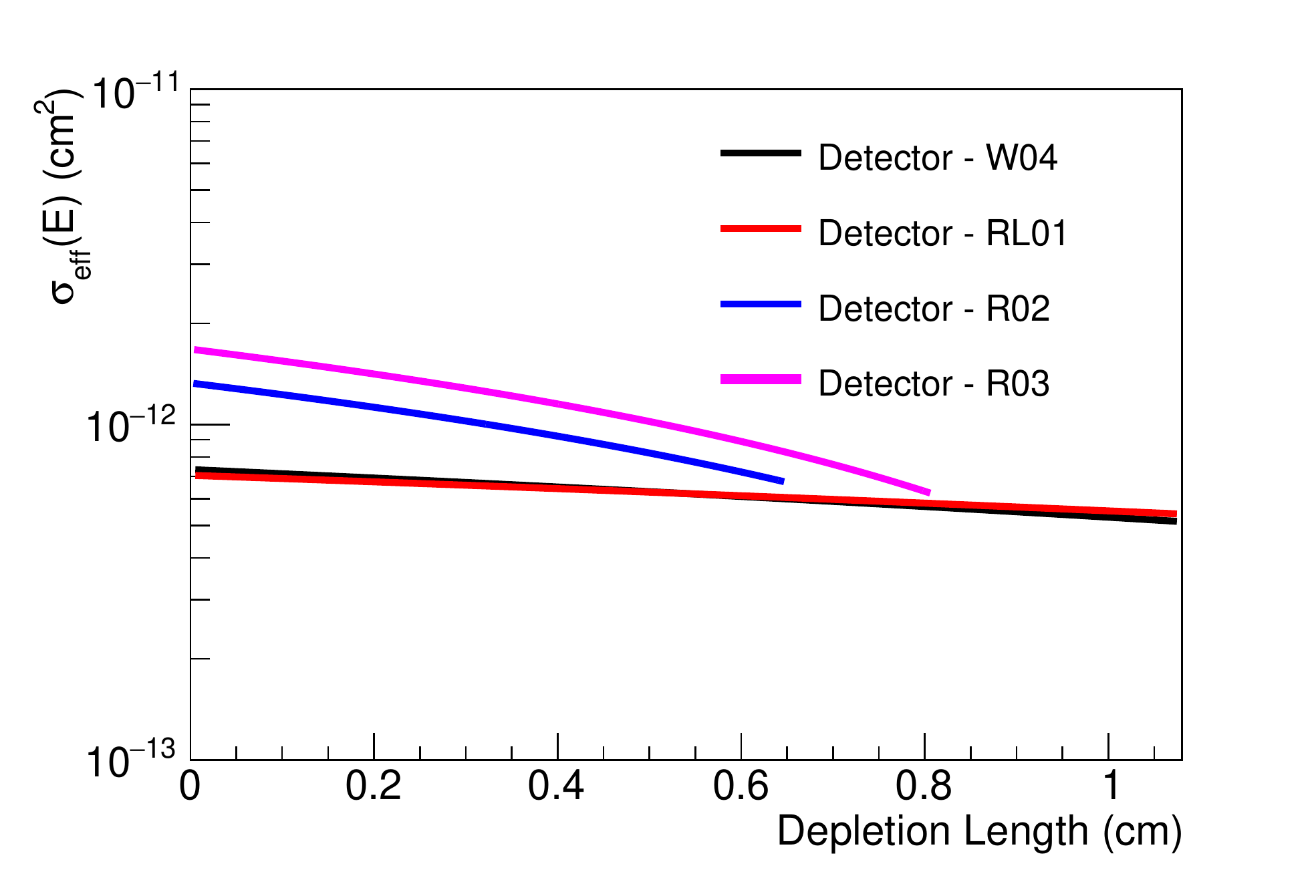}
\caption{\small{Shown is the effective capture cross section distribution for the detectors used in studying electron trapping. 
 }}
\label{fig:crosse}
\end{figure}

\begin{figure}[htb!!!]
\includegraphics[angle=0,width=14.cm] {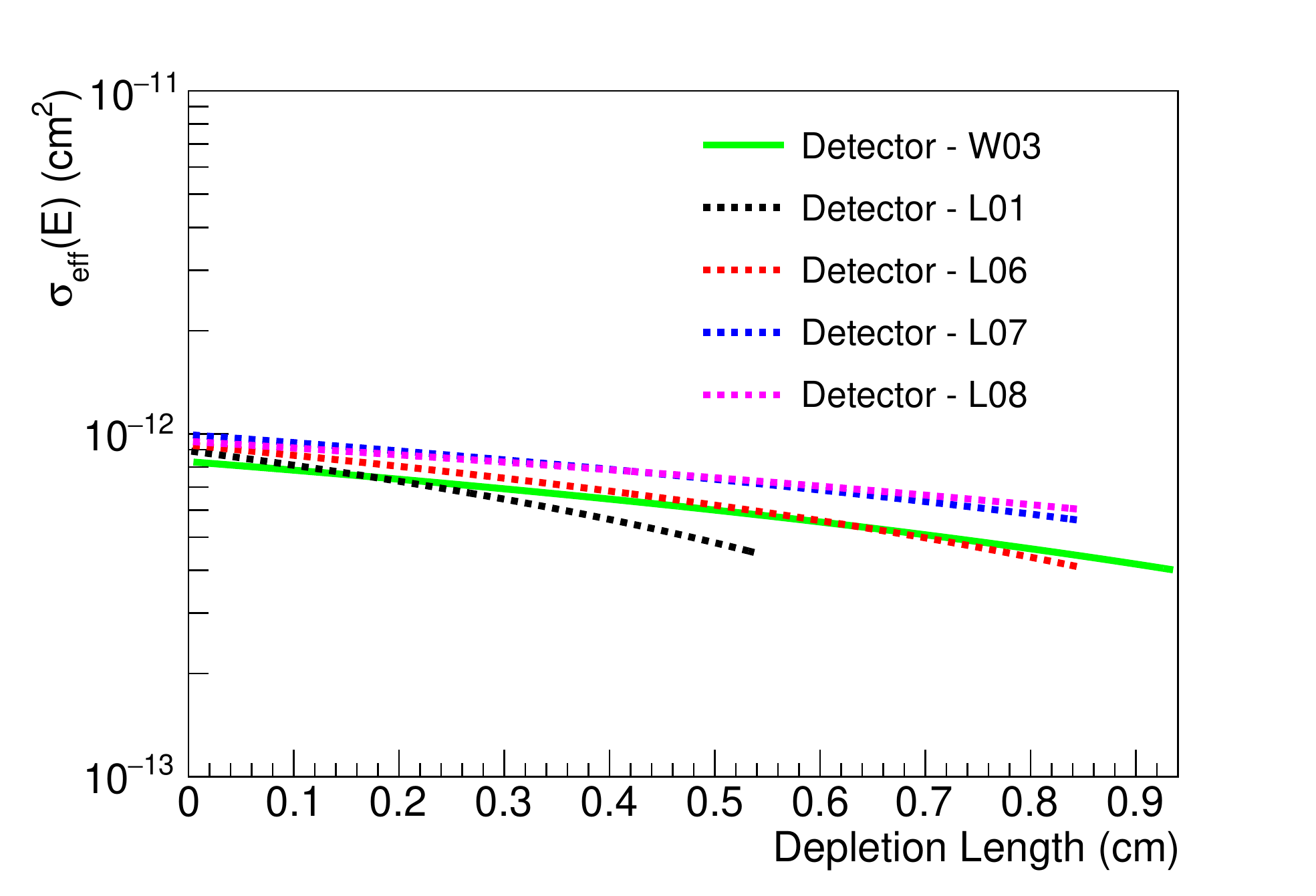}
\caption{\small{Shown is the effective capture cross section distribution for the detectors used in studying hole trapping. 
 }}
\label{fig:crossh}
\end{figure}

One can utilize equations~\ref{velsat} and \ref{scatime} to calculate the mean scattering free path, $\lambda_{c}$ = $v_{d}\times\tau_{c}$, for charge carriers while drifting across the detector. Figures~\ref{fig:meanfe} and \ref{fig:meanfh} show the distributions for all nine detectors. 
Subsequently, the effective capture cross sections can be obtained for all nine detectors utilizing equation~\ref{crosssection}, as shown in Figures~\ref{fig:crosse} and \ref{fig:crossh}.
As a comparison between Figures~\ref{fig:meanfe} and \ref{fig:meanfh}, one can clearly see that the average mean free scattering path of electrons is smaller than that of holes. A smaller mean free scattering path results in a larger capture cross section for electrons. This explains why electron trapping is more severe than hole trapping. It suggests that collecting holes can reduce charge trapping in general. 

\subsection{Calculation of the gain factor due to the impact ionization of impurities}
\begin{figure}[htb!!!]
\includegraphics[angle=0,width=14.cm] {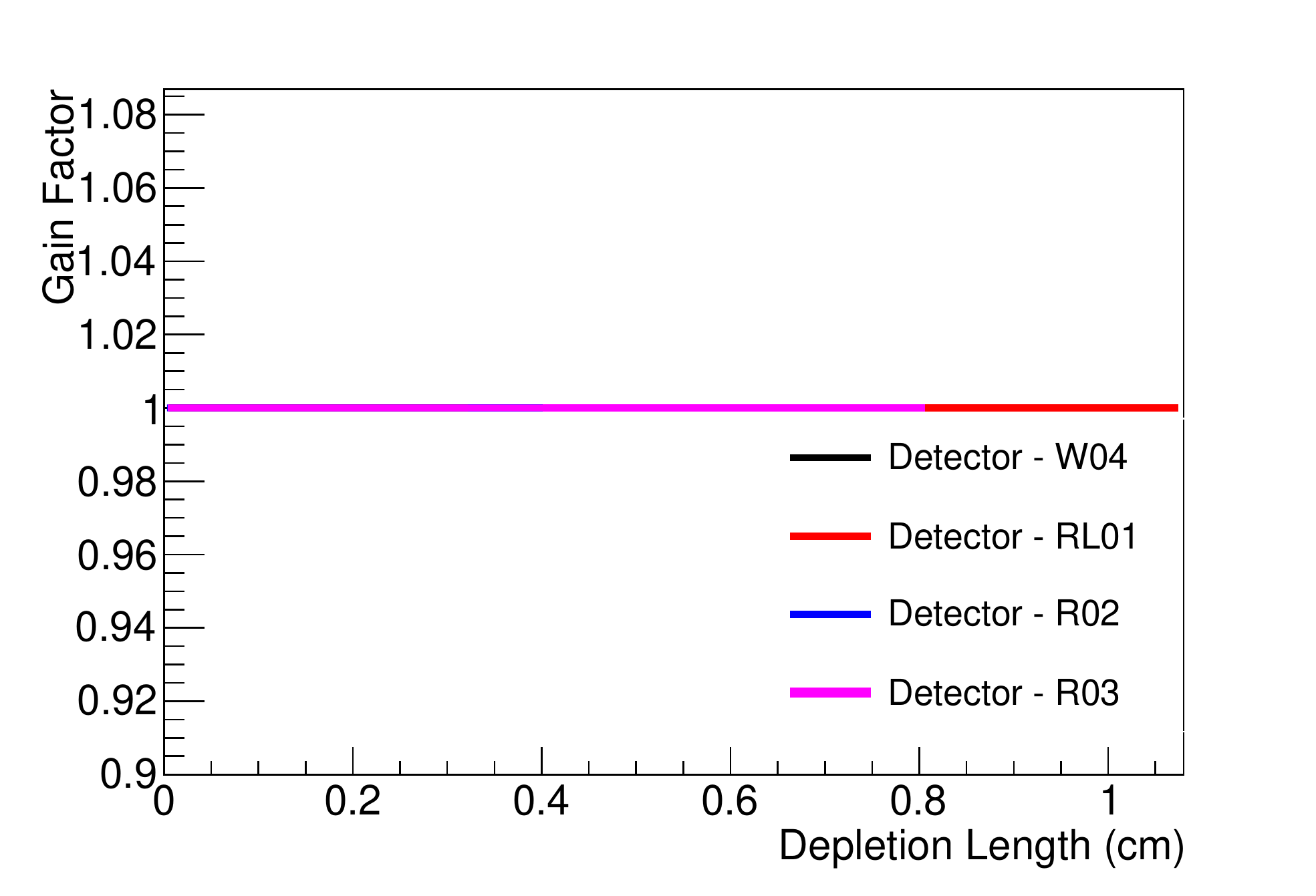}
\caption{\small{Shown is the gain factor distribution for the detectors used in studying electron trapping. Note that the overlapped straight lines for all four detectors are resulted from an unity gain factor, which indicates no impact ionization of impurities for the given operational bias voltage.
 }}
\label{fig:gaine}
\end{figure}

\begin{figure}[htb!!!]
\includegraphics[angle=0,width=14.cm] {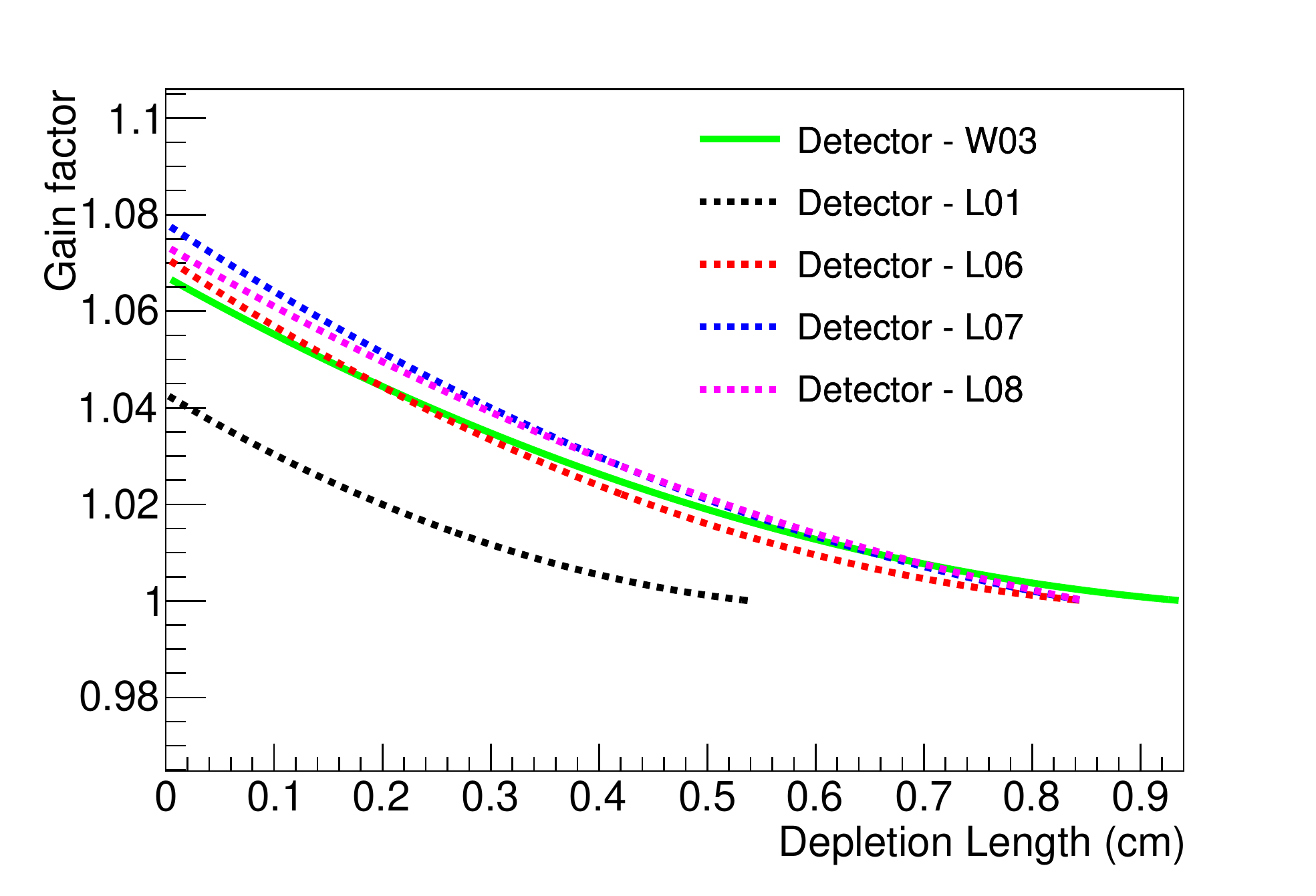}
\caption{\small{Shown is the gain factor distribution for the detectors used in studying hole trapping. 
 }}
\label{fig:gainh}
\end{figure}

The impact ionization of impurities can occur when drifting charge carriers across the detector under a strong electric field. The calculation of the gain factor can be carried out using equation~\ref{gain} for electrons and holes, respectively. Figures~\ref{fig:gaine} and \ref{fig:gainh} show the results for all nine detectors. 

Given the operational bias voltage for the four detectors used in studying electron trapping, no impact ionization of impurities was found. The gain factor for all four detectors is an unity, as shown in Figure~\ref{fig:gaine}. In the case for the five detectors used in studying hole trapping, the visible impact ionization of impurities with a small gain factor from 1.01 to $\sim$1.07 is found in all five detectors. Figure~\ref{fig:gainh} shows the distribution of the gain factor along the drift path. It indicates that a small fraction of charge carriers were produced due to the impact ionization of impurities under the given high electric field when drifting holes across the detectors. 

\subsection{ Determination of charge collection efficiency and the absolute impurity level}
\begin{figure}[htb!!!]
\includegraphics[angle=0,width=14.cm] {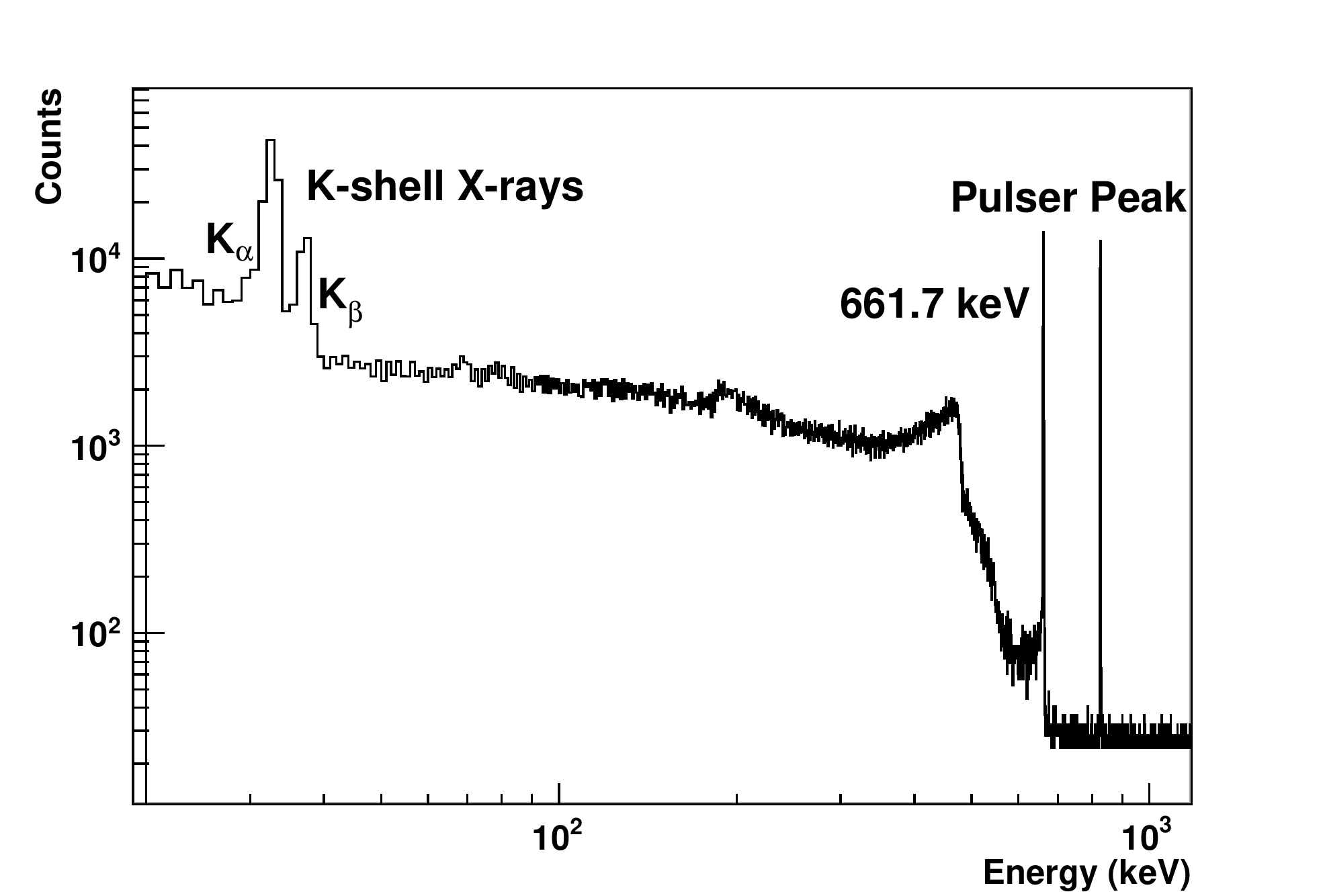}
\caption{\small{Shown is the energy spectrum of $^{137}$Cs measured with one of the USD detectors.  
 }}
\label{fig:cs137}
\end{figure}

Utilizing the measured energy resolution, $\Delta E_{sv}$, for a given X-ray or $\gamma$-ray energy and equation~\ref{14}, we can determine the charge collection efficiency. 
In order to calculate $\Delta E_{sv}$, we first need to determine the values of $\Delta E$ and $\Delta E_{en}$ respectively. This is carried out for analysis with $^{137}$Cs being utilized as the calibration source. Figure~\ref{fig:cs137} shows the energy spectrum of $^{137}$Cs measured with one of the USD detectors. 

\begin{figure}[htb!!!]
\includegraphics[angle=0,width=14.cm] {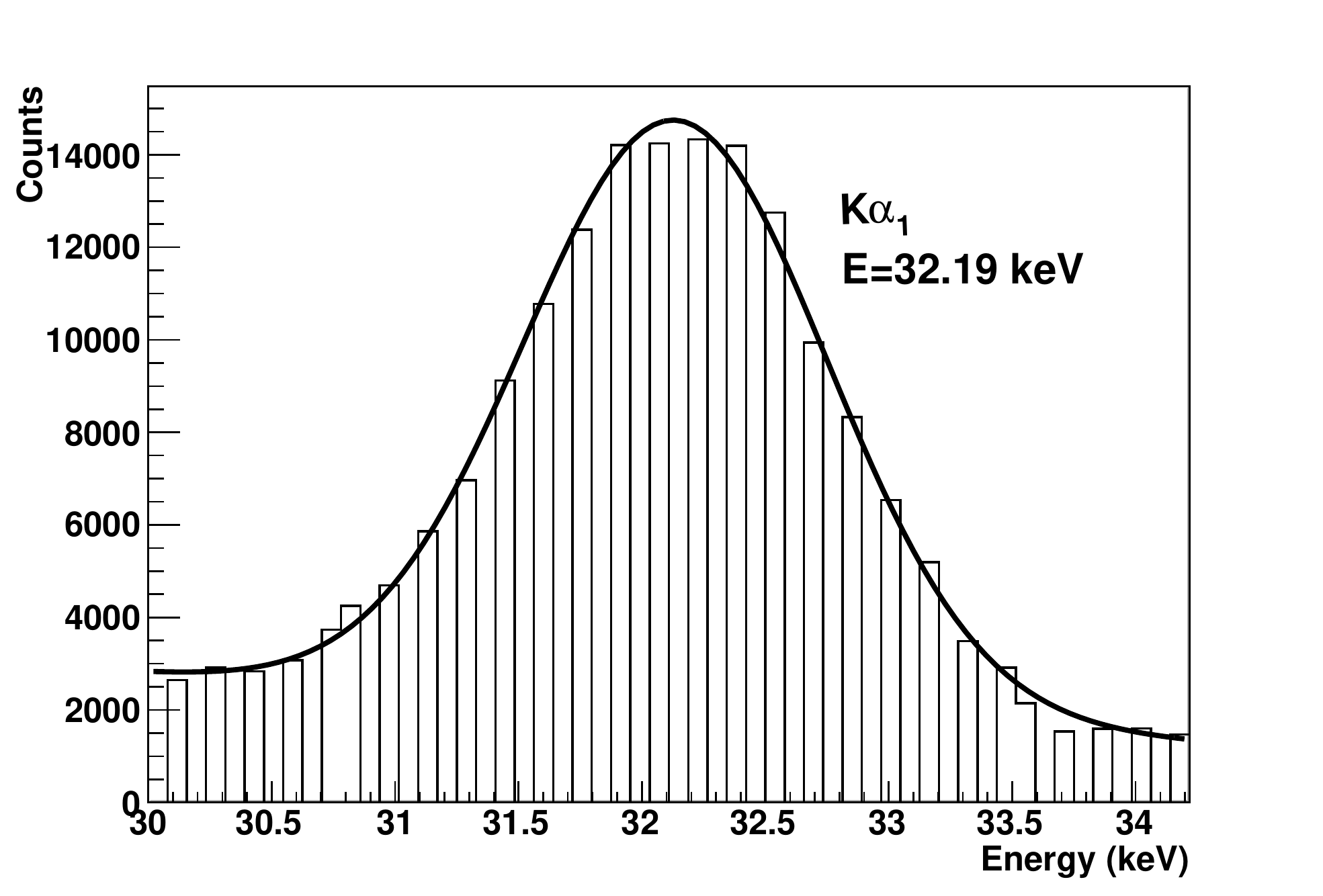}
\caption{\small{Shown is the fitting curve to the measured K $\alpha_{1}$ (32.19 keV) X-ray from $^{137}$Ba produced by internal conversion. The fitted results are: $\chi^{2}/ndf=39.75/53$; Mean = 32.20 keV; $\sigma$ = 0.7205$\pm$0.0023.  
 }}
\label{fig:kalpha}
\end{figure}

Once the energy peaks are calibrated to their corresponding energies, we can calculate the Full Width at Half-Maximum (FWHM),i.e., $\Delta E$ of each peak as it is numerically equal to $2.355 \sigma$. $\sigma$ can be estimated from the coefficients of the Gaussian function used in the corresponding peak fitting. This yields a $\Delta E$ value of the measured total energy resolution (keV) and $\Delta E_{en}$ of the measured electronic noise (keV) for the peak 32.19 keV and the noise peak, respectively. Substituting $\Delta E$ and $\Delta E_{en}$ in equation~\ref{e3}, we obtain $\Delta E_{m}$ (keV) = $\sqrt{\Delta E_{sv}^{2} + \Delta E_{ic}^{2}}$, the convoluted energy resolution, for the 31.19 keV energy peak for all nine detectors, as shown in Table 1. Note that the peak 31.19 keV is the K$_{\alpha_{1}}$ X-ray from $^{137}$Ba produced by internal conversion. We also observed the K$_{\beta}$ X-rays with an average energy of 36.5 keV and 661.7 keV $\gamma$ ray from $^{137}$Cs. All of these three peaks were used for the energy calibration. The peak 32.19 keV is chosen to conduct the trapping analysis, since it guarantees the creation of e-h pairs at the surface of the detector for each configuration shown in Figure~\ref{fig:setup}. As an example, Figure~\ref{fig:kalpha} displays a fitting curve for the K$\alpha_{1}$ X-ray measured with one of the USD detectors. 

Putting the calculated gain factor ($g_{c}$), the Fano factor (0.106), and the measured energy resolution ($E_{m}$) into equation~\ref{14}, the average charge collection efficiency ($\bar{\varepsilon}_{h}$) is obtained. Applying $\bar{\varepsilon}_{h}$ to equation~\ref{e100}, one can determined the average trapping length ($\bar{\lambda}_{th}$) for each detector. Subsequently, utilizing the calculated effective cross section ($\bar{\sigma}_{eff}(E)\times<v_{tot}>/<v_{d}>$) and the average trapping length with equation~\ref{trappinglength}, we can determine the absolute impurity level for all nine detectors. As a summary, the detector name, measured total energy resolution, measured noise width, and determined net impurity level for each detector are given in Table~\ref{tab:t2}. 

\begin{table}[h]
\centering
\caption{An overview of the results of nine detectors used in this analysis. $\Delta E$ stands for the measured total energy resolution. $\Delta E_{en}$ is the measured full width at half maximum for the electronic noise. $\Delta E_{m}$ = $\sqrt{\Delta E^2 - \Delta E_{en}^2}$. $\bar{\varepsilon}$ is the average charge collection efficiency. $\bar{\lambda_{th}}$ is the average trapping length. $N_A+N_D$ is the absolute impurity.  Note that the uncertainties associated with parameters in Table~\ref{tab:t2} are discussed in the text. }
\label{tab:t2}
\begin{tabular}{|c|c|c|c|c|c|c|}
\hline \hline
Detector & $\Delta E$& $\Delta E_{en}$ & $\Delta E_{m}$ & $\bar{\varepsilon}_{h}$ & $\bar{\lambda}_{th}$&$N_A+N_D$\\ 
& keV & keV & keV &(\%)&(cm)&10$^{11}$(cm$^{-3}$)\\ \hline\hline
 W04&2.4354$\pm$0.153&2.41$\pm$0.15&0.351$^{+0.035}_{-0.042}$&45.55$^{+13.19}_{-7.95}$&0.58$^{+0.31}_{-0.14}$&14.10$^{+0.55}_{-0.61}$\\ \hline
 W03&2.115$\pm$0.041&2.10$\pm$0.04&0.251$^{+0.003}_{-0.013}$&88.86$^{+10.14}_{-1.79}$&3.90$^{+20.15}_{-0.80}$&1.85$^{+0.61}_{-1.03}$\\ \hline
 RL01 &1.965$\pm$0.102&1.93$\pm$0.10&0.368$^{+0.018}_{-0.020}$&41.29$^{+4.83}_{-3.70}$&0.51$^{+0.09}_{-0.05}$&14.70$^{+0.31}_{-0.38}$\\ \hline
 R02&1.697$\pm$0.092&1.67$\pm$0.09&0.302$^{+0.017}_{-0.018}$&61.35$^{+7.84}_{-5.83}$&0.62$^{+0.21}_{-0.15}$&8.55$^{+0.29}_{-0.30}$\\ \hline
 R03 &1.288$\pm$0.085&1.23$\pm$0.06&0.382$^{+0.002}_{-0.005}$&38.47$^{+0.91}_{-0.44}$&0.35$^{+0.01}_{-0.02}$&13.50$^{+0.86}_{-0.38}$\\ \hline
 L01&0.999$\pm$0.011&0.97$\pm$0.01&0.238$^{+0.035}_{-0.042}$&99.20$^{+0.72}_{-0.70}$&35.50$^{+30.50}_{-0.12}$&0.22$^{+0.21}_{-0.31}$\\ \hline
 L06&1.702$\pm$0.092&1.67$\pm$0.09&0.328$^{+0.009}_{-0.021}$&52.20$^{+7.31}_{-3.01}$&0.58$^{+0.17}_{-0.07}$ &13.40$^{+1.90}_{-2.12}$\\ \hline
 L07&1.214$\pm$0.045&1.19$\pm$0.04&0.240$^{+0.002}_{-0.004}$&97.60$^{+2.30}_{-2.30}$&17.50$^{+60.51}_{-8.51}$ &0.39$^{+0.22}_{-0.32}$\\ \hline
 L08&1.057$\pm$0.021&1.03$\pm$0.02&0.239$^{+0.003}_{-0.002}$&97.87$^{+2.10}_{-2.10}$&21.50$^{+55.65}_{-12.10}$&0.32$^{+0.42}_{-0.24}$\\ \hline \hline
\end{tabular}
\end{table}

The uncertainties on $<v_d>$ and $<v_{tot}$ are evaluated to be less than 8\%, which is obtained using the uncertainty on the Hall mobility (10\%) from the IEEE standard when compared to the Hall Effective measurements performed in our lab and the uncertainty of $v_{sat}$ (4\%) using equation~\ref{satvel} when compared to the experimental result from W. M\"{o}noh~\cite{wmo}. 
Since there are multiple parameters used in calculating the gain factor ($g_c$) and the Fano factor, which are used to determine the average charge collection efficiency ($\bar{\varepsilon}_{h}$) and hence the average trapping length ($\bar{\lambda}_{th}$), as well as the effective cross section ($\bar{\sigma}_{eff}(E)\times<v_{tot}>/<v_{d}>$), the uncertainty is evaluated based on our best knowledge. First, the uncertainty associated with calculating $g_c$ is evaluated using each parameter involved in equation~\ref{gain}, which the uncertainty of the electric field is less than 1\% and the uncertainty on the mean free path ($\lambda_c$) is evaluated using the uncertainty of $v_d$, the uncertainty of $v_{sat}$, and the small uncertainty of the electric field by adding them in quadratic form, since they are independent from each other. The resulted uncertainty of $\lambda_c$ is 11.7\%. The uncertainty of $\lambda_R$ is mainly from the uncertainty of $\tau_{ph}$, which is from the uncertainty of the Hall mobility (10\%) stated above.  The other parameters in equation~\ref{gain} have standard values, which are widely used in the field, and hence no uncertainty assigned to those parameters. Therefore, the uncertainty of $g_c$ is obtained to be 0.6\% by propagating the uncertainty of the electric field and the uncertainty of $\lambda_c$ in equation~\ref{gain}.  Secondly, the uncertainty on the effective cross section is dominated by the uncertainty of $\lambda_c$, since the uncertainty on the ratio of  $<v_{tot}>/<v_{d}>$ is less than 1\%. Thus, the uncertainty of the effective cross section is obtained to be 11.7\%. It is worth mentioning that the drift velocity $<v_d>$ used in equation~\ref{trappinglength} is the averaged group velocity, which will not be influenced by scattering of individual charge carrier off acceptors or donors significantly. This treatment has been widely used in the SuperCDMS experiment~\cite{sundqvist, phipps}. Similarly, scattering off displacement defects will also not influence $<v_d>$ significantly. Thus, we don't consider these two scattering processes as the possible sources of uncertainty associated with $<v_d>$.  

The uncertainty associated with $\Delta E$ is the total error determined through the fitting method described earlier. Since the data was taken for 30 minutes with a strong $^{137}$Cs source, the statistical error is at a level of $\sim$1\%. Therefore, the total error from the fitting is dominated by the contamination from the K$\alpha_{2}$ X-ray, which has an energy of 31.82 keV with a 54\% of intensity relative to the K$\alpha_{1}$ X-ray. The main uncertainty originates from the variation of electronic noise. This can be also seen from the measured $\Delta E_{en}$ with different detectors. Since the measurements were made at different time, the width of noise peak changed from about $\sim$1 keV to over 2 keV in a year. To minimize the variation of electronic noise, the data was only taken for 30 minutes. The noise peak was taken before and after a measurement. The uncertainty associated with $\Delta E_{en}$ is the variation of noise within 30 minutes. The calculated uncertainty for $\Delta E_{m}$ is mainly from the propagation of the uncertainty of electronic noise.  

The uncertainties associated with the average charge collection efficiency and the average trapping length are mainly from the uncertainty of the gain factor, the uncertainty of the effective cross section, and the uncertainty of electronic noise. It is clear that a stable and a smaller electronic noise is preferred for studying charge trapping. Note that the uncertainties quoted for the average trapping lengths, which correspond to a charge collection of $\sim$90\% or above, are not very meaningful, since the trapping length is insensitive to the range of charge collection efficiency $\geq$ $\sim$90\%. In this range, as described by equation~\ref{e100}, a small change of charge collection efficiency requires a large change of charge trapping length. The uncertainty on the charge collection efficiency due to the variation of the drifting length from the creation of charge carriers is less than 1\%, since the mean free path of 32.19 keV X-ray in Ge is 0.016 cm, which is much smaller than the detector depletion thickness.

With the measured energy resolution for each detector, we calculate the average charge collection efficiency, as shown in Table~\ref{tab:t2}. 
Once we know the average charge collection efficiency for each detector, the average trapping length can be calculated using equation~\ref{e100} and the results are shown in Table~\ref{tab:t2}. Subsequently, equation~\ref{trappinglength} can be used to calculate the absolute impurity concentration for each detector. The results are shown in Table~\ref{tab:t2}.

In theory, it is expected that almost all the charge carriers for a given signal can be collected for the given detector. However, it has been shown above that this is not always the case as the charge collection efficiency is not always 100\% for the given detectors.
This indicates that the trapping process for electrons and holes must be of the forms: $e^- + D^+ \longrightarrow D^0$ and $h^{+} + A^{-} \longrightarrow A^{0}$ (assuming that the trapping centers are predominantly singly charged). The formation of the neutral trapping center is relatively stable due to the fact that phonon excitation at 77 K is not energetically high enough to re-release the $e^-/h^+$ trapped by these impurity centers.  

\subsection{Discussion of implications} 

As can be seen in Table~\ref{tab:t2}, the absolute impurity levels are in the range of (8.6$-$14.7)$\times$10$^{11}$/cm$^{3}$ for the four detectors that were used to study electron trapping and (0.32 $-$13.4)$\times$10$^{11}$/cm$^{3}$ for the five detectors used in studying hole trapping. If one compares the absolute impurity levels ($N_A+N_{D}$) to the net impurity level ($|N_A - N_D|$) listed in Table~\ref{tab:t2}, we can conclude that the absolute impurity levels can be a factor of a few to 100 higher than the net impurity levels. For example, the absolute impurity level in the detector L08 is only a factor of less than 2 higher than the net impurity level. This indicates that the n-type impurity in L08 is smaller than the p-type impurity and L08 is quite pure. On the other hand, the detectors W04 and RL01 have a net impurity level of $\sim$10$^{9}$/cm$^{3}$, the absolute impurity levels are in the range of $\sim$10$^{12}$/cm$^{3}$. This implies that the crystals that were used to fabricate detectors W04 and RL01 were highly compensated. To fully deplete a Ge detector, only the net impurity level is critical. In general, as long as a detector can be fully depleted,  it can be a good detector in terms of detecting $\gamma$-ray radiation. However, the energy resolution can be affected by the absolute impurity level due to the energy loss through charge carrier capture by p-type or n-type impurities. This is particularly important for the rare-event physics experiments that require extremely good energy resolution. 

\begin{figure}[htb!!!]
\includegraphics[angle=0,width=14.cm] {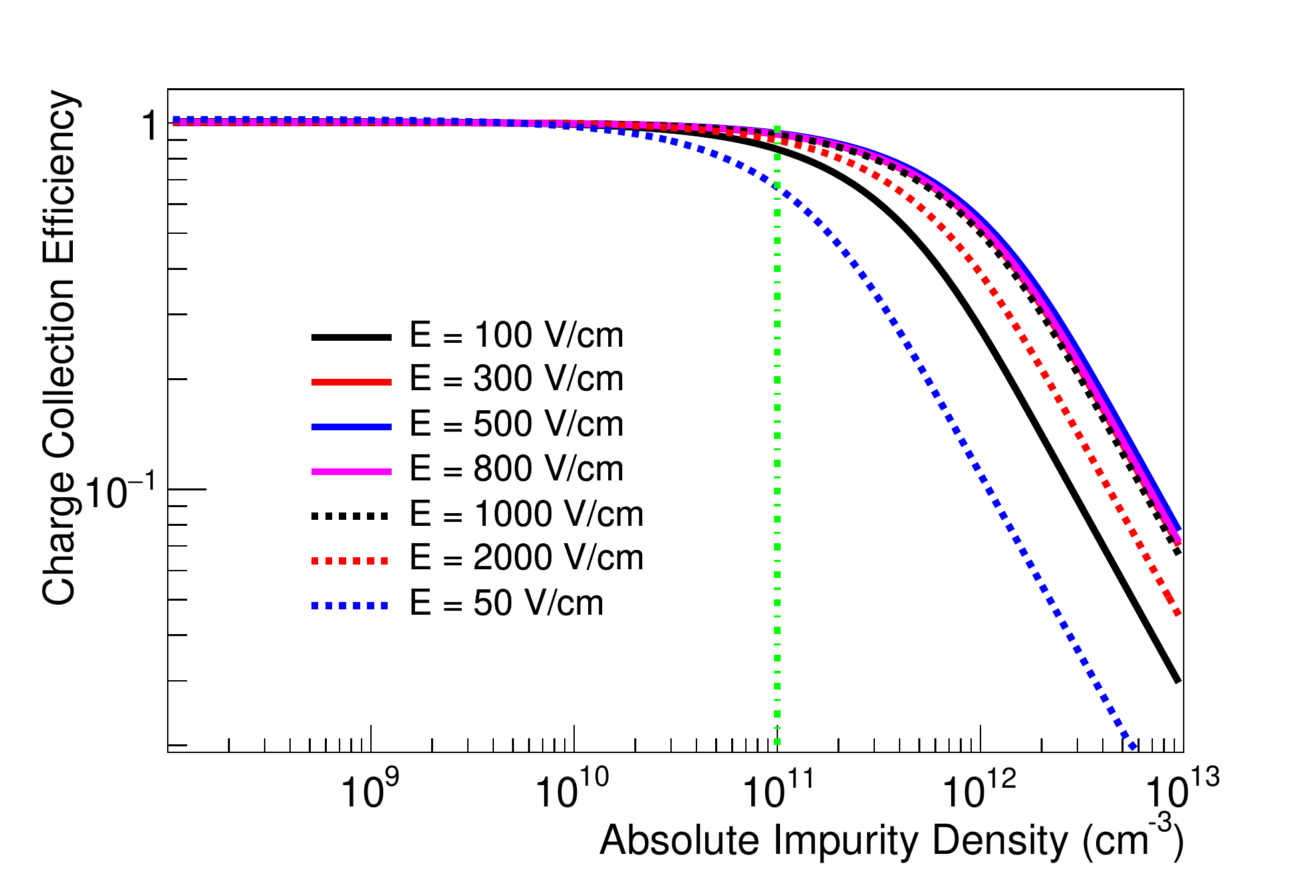}
\caption{\small{Shown is the charge collection efficiency as a function of the absolute impurity while collecting electrons. 
 }}
\label{fig:chargece}
\end{figure}

\begin{figure}[htb!!!]
\includegraphics[angle=0,width=14.cm] {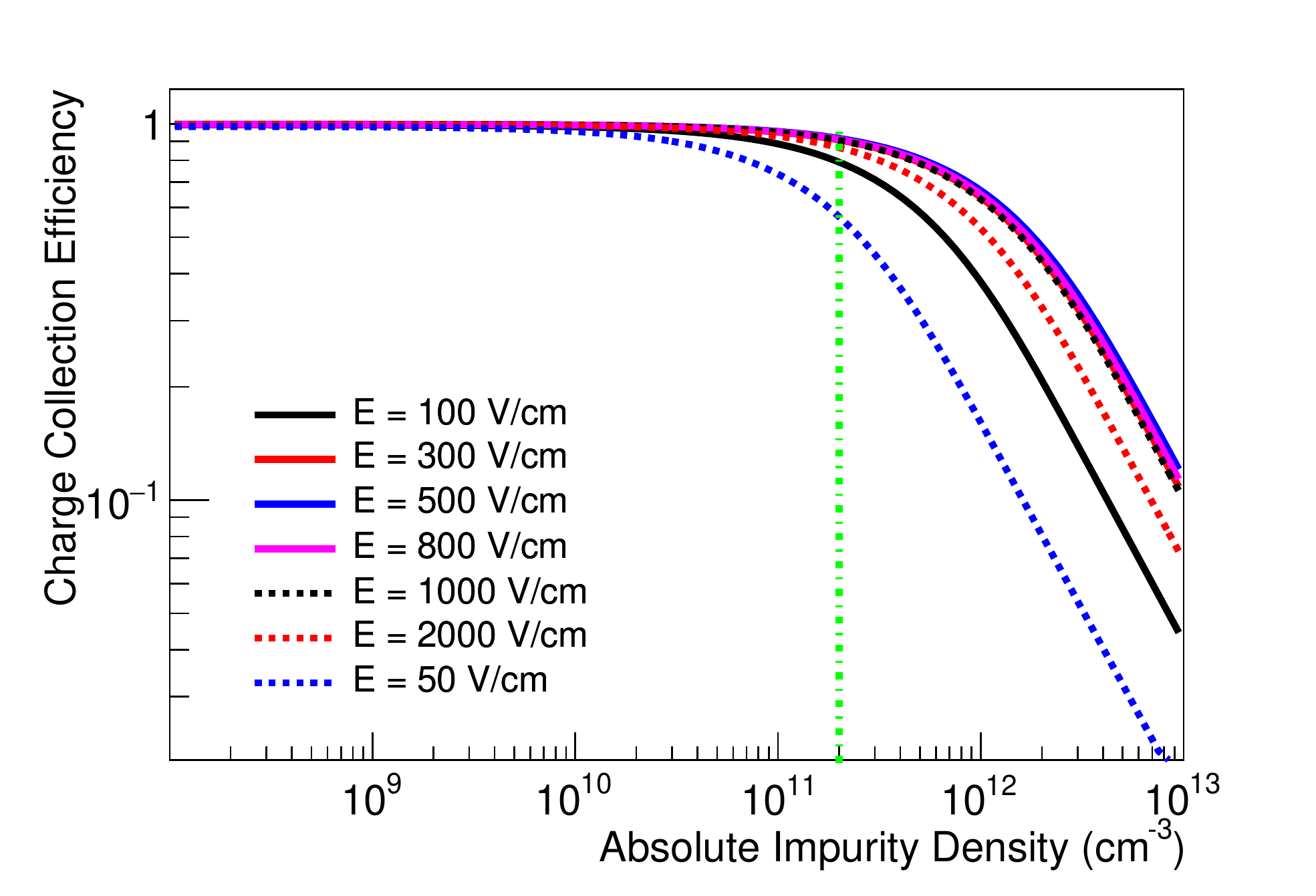}
\caption{\small{Shown is the charge collection efficiency as a function of impurity while collecting holes. 
 }}
\label{fig:chargech}
\end{figure}

The detectors fabricated by commercial companies usually come with information about the net impurity level, which guarantees the full depletion of the detectors with the recommended bias voltage. However, charge trapping is related to either p-type or n-type impurities, depending on the choice of electron collection or hole collection. Therefore, knowing the impact of the absolute impurity on the energy resolution will allow us to understand the detector performance better in terms of charge trapping. Utilizing equation~\ref{e100}, we can estimate the charge collection efficiency as a function of the absolute impurity for a full depleted detector with a thickness of 3 cm and a net impurity level of 5$\times$10$^{9}$/cm$^{3}$. Figures~\ref{fig:chargece} and \ref{fig:chargech} show the results, which suggest that a charge collection efficiency of $>$ 90\% can be achieved in collecting electrons when the absolute impurity level is less than 1.0$\times$10$^{11}$/cm$^3$. A similar efficiency can be achieved in collecting holes when the absolute impurity level is less than 2.0$\times$10$^{11}$/cm$^3$. This difference in the requirement of the absolute impurity level corresponds to a larger cross section of electrons when comparing to holes. 


\begin{figure}[htb!!!]
\includegraphics[angle=0,width=14.cm] {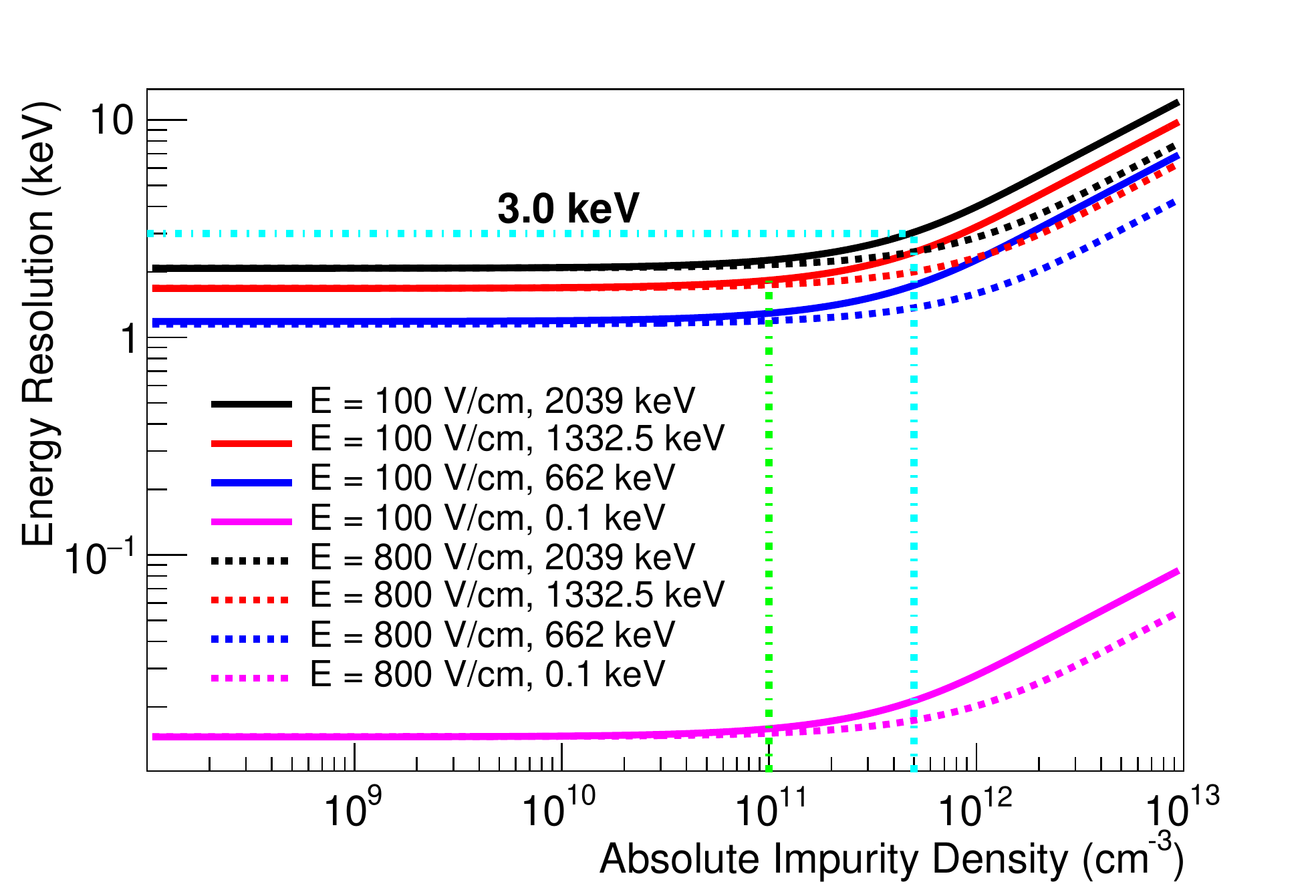}
\caption{\small{Shown is the energy resolution as a function of impurity while collecting electrons. 
 }}
\label{fig:erue}
\end{figure}

\begin{figure}[htb!!!]
\includegraphics[angle=0,width=14.cm] {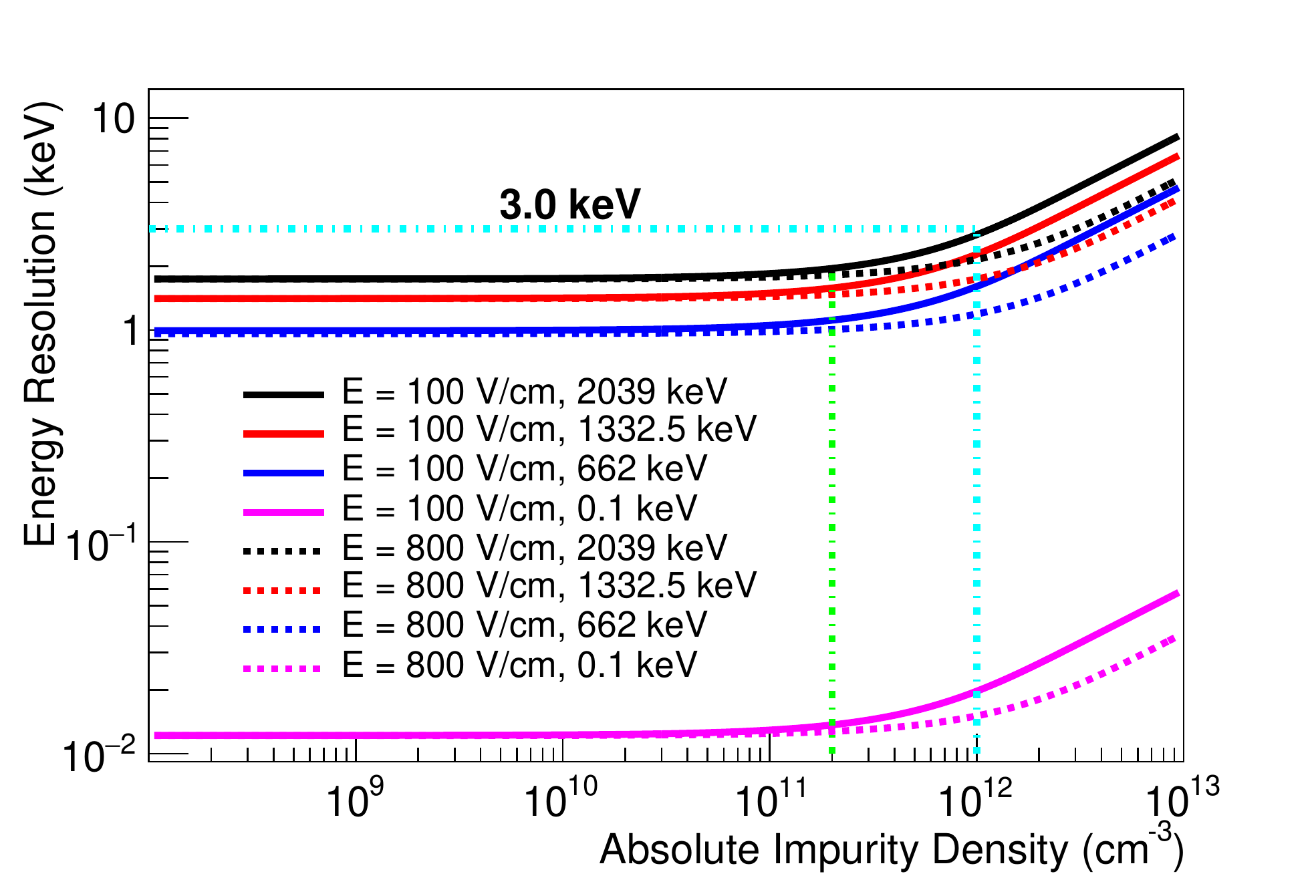}
\caption{\small{Shown is the energy resolution as a function of impurity while collecting holes. 
 }}
\label{fig:eruh}
\end{figure}

One can translate the charge collection efficiency into the energy resolution using equation~\ref{14}. We display the energy resolution as a function of the absolute impurity in Figures~\ref{fig:erue} and ~\ref{fig:eruh}. Interestingly, we notice that the best energy resolution can be achieved for several energy lines, such as 2039 keV for 0$\nu\beta\beta$ decay using $^{76}$Ge, three calibration $\gamma$-ray energies from $^{60}$Co and $^{137}$Cs, as well as 100 eV for low mass dark matter searches, when the absolute impurity level is less than 1.0$\times$10$^{11}$/cm$^3$ in collecting electrons and 2.0$\times$10$^{11}$/cm$^3$ in collecting holes. This is commonly achievable from commercial companies and from the USD-grown crystals, as demonstrated by detectors L01, L07, and L08.


One must point out that the best energy resolution of 2 keV at 2039 keV can be achieved with the absolute impurity level of $<$2.0$\times$10$^{11}$/cm$^3$ in collecting holes and $<$ 1.0$\times$10$^{11}$/cm$^3$ when collecting electrons. The energy resolution of 3 keV at 2039 keV would correspond to an absolute impurity level of $\sim$1$\times$10$^{12}$/cm$^3$, which can usually be prevented for the crystals from commercial companies. This indicates that the measured energy resolution of $\sim$3 keV in the energy region of 2000 keV~\cite{5,6} can be further improved by reducing the electronic noise from the data acquisition system. Similar conclusion can be made for low-energy at 100 eV. The detector technology is able to provide the needed energy resolution if the electronic noise can be brought under control. Note that charge trapping depends also on the applied electric field. If a significant degradation of energy resolution is observed, it usually indicates that the detector has not only a high absolute impurity level, but also a weak electric field in some regions.   

\section{Conclusion}
\label{sec:conl}
We present a comprehensive study of charge trapping utilizing nine planar detectors fabricated at USD. Among them, four detectors were used to study electron trapping and five detectors were used to study hole trapping. All detectors were over-biased to guarantee sufficient electric field in order to only study deep level charge trapping - charge carrier captured by impurity centers. Since the detectors are small, the assumption of the uniform impurity distribution is valid. This allows us to utilize the well-established models to obtain the charge collection efficiency for all detectors. Furthermore, we can determine the average trapping length for each detector using the measured energy resolution. Once the trapping length is known, we can determine the absolute impurity level for each detector, since the net impurity level is measured through the full depletion voltage and the detector thickness. Thus, we can correlate the energy resolution with the absolute impurity level and investigate the impact of the absolute impurity on the energy resolution. This sheds light on the large size detectors for 0$\nu\beta\beta$ decay and dark matter searches. We conclude that the best energy resolution can be achieved when the absolute impurity is less than 2.0$\times$10$^{11}$/cm$^3$ in collecting holes and 1.0$\times$10$^{11}$/cm$^3$ in collecting electrons. These level of impurity is commonly achievable by the current crystal growth and detector fabrication technology. One has to make clear is that the net impurity can be in the range of $\sim$10$^{9}$/cm$^3$ while the absolute impurity can be in the range of $\sim$10$^{12}$/cm$^3$, which can introduce significant charge trapping and decrease the energy resolution. Therefore, the net impurity level is only critical to fully deplete the detector. It is the absolute impurity level that contributes to the energy resolution. Thus, it is important to emphasize the criteria of the detector requirement by including the absolute impurity level. As long as the net impurity level allows the detector to be fully depleted with sufficient bias voltage, achieving the excellent energy resolution will depend on the absolute impurity level and its distribution inside the detector when operating under a field of more than 100 V/cm across the entire detector. 

\section*{Acknowledgement} 
The authors would like to thank Mark Amman for his instruction on fabricating planar detectors and Christina Keller for a careful reading of this manuscript. We would also like to thank the PIRE-GEMADARC collaborators John Wilkerson, Iris Abt, Nader Mirabolfathi, Wenqin Xu, and Nicholas Mast for useful discussions and comments.  In addition, we would like to thank the Nuclear Science Division at Lawrence Berkeley National Laboratory for providing us a testing cryostat. This work was supported in part by NSF NSF OISE 1743790, NSF PHYS 1902577, NSF OIA 1738695, DOE grant DE-FG02-10ER46709, DE-SC0004768, the Office of Research at the University of South Dakota and a research center supported by the State of South Dakota.

\end{document}